\documentclass{aastex62}

\usepackage{array}
\usepackage{multirow}
\usepackage[utf8]{inputenc}

\newcommand{\msun}{\ensuremath{M_\sun}}

\graphicspath{{./}{figures/}}

\newcommand{\beq}{\begin{equation}}
\newcommand{\eeq}{\end{equation}}

\shorttitle{Orbits of K Dwarfs}
\shortauthors{Horch et al.}

\begin{document}
\title{Observations with the Differential Speckle Survey Instrument. 
X. Preliminary Orbits of K Dwarf Binaries and Other Stars}

\correspondingauthor{Elliott Horch}
\email{horche2@southernct.edu}

\author{Elliott P. Horch}
\altaffiliation{Adjunct Astronomer, Lowell Observatory; Visiting Astronomer, Kitt Peak National Observatory;
Visiting Astronomer, Gemini Observatory}
\affiliation{Department of Physics, Southern Connecticut State University, 501 Crescent Street, New Haven, CT 06515, USA}

\author{Kyle G. Broderick}
\altaffiliation{Current Address: Aerospace/Hydrospace Engineering \& Physical Sciences, Fairchild Wheeler Interdistrict Magnet Campus, 840 Old Town Road, Bridgeport, CT, 06606}
\affiliation{Department of Physics, Southern Connecticut State University, 501 Crescent Street, New Haven, CT 06515, USA}

\author{Dana I. Casetti-Dinescu}
\affiliation{Department of Physics, Southern Connecticut State University, 501 Crescent Street, New Haven, CT 06515, USA}

\author{Todd J. Henry}
\affiliation{RECONS Institute, Chambersburg, PA 17201, USA}



\author{Francis C. Fekel}
\altaffiliation{Visiting Astronomer, Kitt Peak National Observatory}
\affiliation{Center of Excellence in Information
Systems, Tennessee State University, 3500 John A. Merritt
Boulevard, Box 9501, Nashville, TN 37209, USA}

\author{Matthew W. Muterspaugh}
\affiliation{Columbia State Community College, 1665 Hampshire Pike, Columbia, TN 38401, USA}

\author{Daryl W. Willmarth}
\affiliation{NSF's National Optical-Infrared Research Laboratory, 950 N. Cherry Avenue, Tucson, 
AZ 85719, USA}

\author{Jennifer G. Winters}
\altaffiliation{Visiting Astronomer, Gemini Observatory}
\affiliation{Center for Astrophysics, Harvard \& Smithsonian, 60 Garden Street, Cambridge, MA 02138, USA}

\author{Gerard T. van Belle}
\altaffiliation{Visiting Astronomer, Kitt Peak National Observatory}
\affiliation{Lowell Observatory, 1400 West Mars Hill Road, Flagstaff, AZ 86001}

\author{Catherine A. Clark}
\altaffiliation{Visiting Astronomer, Kitt Peak National Observatory}
\affiliation{Lowell Observatory, 1400 West Mars Hill Road, Flagstaff, AZ 86001}
\affiliation{Department of Astronomy and Planetary Science, Northern Arizona University, Flagstaff, AZ 86001, USA}

\author{Mark E. Everett}
\affiliation{NSF's National Optical-Infrared Research Laboratory, 950 N. Cherry Avenue, Tucson, 
AZ 85719, USA}




\begin{abstract}
This paper details speckle observations of binary stars taken at the Lowell 
Discovery Telescope, the WIYN telescope, and the Gemini telescopes between 2016 January 
and 2019 September. The observations 
taken at Gemini and Lowell were done with the 
Differential Speckle Survey Instrument (DSSI), and those done at WIYN were taken with the 
successor instrument to DSSI at that site, the NN-EXPLORE Exoplanet
Star and Speckle Imager (NESSI). In total, we present 378 observations of
178 systems and we show that
the uncertainty in the measurement precision for the combined data set
is $\sim$2 mas in separation, $\sim$1-2 degrees in 
position angle depending on the separation, and $\sim$0.1 magnitudes in magnitude difference. Together
with data already in the literature, these new results permit 25 visual orbits and
one spectroscopic-visual orbit to be calculated for the first time. 
 In the case of the spectroscopic-visual 
 analysis, which is done on the trinary star HD 173093, we calculate masses with precision 
 of better than 1\% for
 all three stars in that system.
Twenty-one of the visual orbits calculated have a K dwarf as the primary star;
we add these to the known orbits of K dwarf primary stars and discuss the 
basic orbital properties of these 
stars at this stage. Although incomplete, the data that exist so far indicate that binaries with K dwarf
primaries tend not to have low-eccentricity orbits at separations of one to a few tens of AU, that is, on
solar-system scales.
\end{abstract}


\section{Introduction}

While there have been comprehensive multiplicity studies of G and M dwarfs, K dwarfs have remained
somewhat neglected observationally. \citet{duc13} gave a thorough review of stellar multiplicity from 
formation mechanisms to observational statistics, yet K dwarfs were not discussed, due to the 
relative lack 
of data available at the time compared with other spectral types. 
Yet, what we know observationally about G and M dwarfs motivates 
a careful study of K dwarfs even from the perspective of the most basic statistics. The G dwarf 
multiplicity rate has been well-established to be about 50\% non-single \citep{duq91,rag10},
and more recent work by \citet{win19} indicates that M dwarfs have a much lower multiplicity rate,
roughly half that of G dwarfs, so that a study of K dwarfs will allow us to see at what mass this transition
occurs, and how sharp it is.
K dwarf stars are also important for exoplanet
research, as they are approximate analogues to our Sun in important
ways.  Because K dwarfs are roughly twice as common as G dwarfs, in a
volume-limited sample they provide a larger set of stars that can be
analyzed in statistical studies to reveal clues about star and planet
formation history.  In addition, large subsets of stars can be sensibly
sub-divided to learn more about parameters which may influence the
formation outcomes, such as age, metallicity, and magnetic properties.
Unlike M dwarfs, they are relatively
bright and easily observed in a survey capacity at 4-m class telescopes. 
At the same time, they represent a more favorable situation
in terms of observational ease in detecting transit signals compared to G dwarfs, and offer shorter orbital 
periods for exoplanets that are in their habitable zones.

For these reasons, since 2015, our collaboration has been working to provide the same 
kind of volume-complete, unbiased statistics for K dwarfs that exist for the other lower-mass spectral 
types on the Main Sequence. This is being accomplished with a combination of three separate surveys
that together span the complete range of separations for gravitationally-bound companions. 
First, we are surveying nearby
K stars with spectroscopy, identifying spectroscopic binaries \citep{par21}. Second, we are 
completing a wide-field survey
using catalogs such as the {\it Gaia} DR2 \citep{gai18} and EDR3 \citep{gai20}. 
Third, to bridge the separation regimes between
these two surveys, we are observing the same list of stars using speckle imaging at a combination
of facilities, including both Gemini telescopes, the WIYN telescope, and the Lowell Discovery Telescope
(LDT). This paper presents results from the speckle survey on measurements of the companions of 
K dwarf stars and some other targets of interest, and uses that information together with previous measures 
to derive visual orbital elements for 26 systems in all, 21 of which are K dwarfs. 
This brings the total number of visual orbits for systems that have a K dwarf primary to 246; thus, this 
paper increases the sample by approximately 9\%.

\section{Observations and Data Reduction}

The results presented here are derived from observing runs which took place at the LDT,
WIYN, and the two Gemini telescopes from 2016 through 2019. 
The Differential Speckle Survey
Instrument (DSSI) was used at the LDT for the majority of the observations \citep{hor09}. We then 
supplemented this data set with a run at the WIYN telescope with the NESSI speckle
camera \citep{sco18} in 2019 two nights of which was provided via Director's Discretionary Time. A few 
further observations taken at Gemini-S in 2017 June and at Gemini-N in 2016 January are also 
included that bear directly on the orbit 
calculations presented later in this paper. We have obtained a large number of observations for the
program from both the WIYN and Gemini telescopes; some other Gemini measures are included in 
\citet{nus21}, but the majority of our Gemini and WIYN queue observations obtained to date
will be published in a future paper. In the current paper, we seek to focus primarily on objects where 
some previous observations exist, so that orbits
can be calculated either now or in the near future. A short table of the observing runs and instrument 
combinations in the current data set is given in Table 1.

\begin{deluxetable}{rlllc}
\tablewidth{0pt}
\tablenum{1}
\tablecaption{Observing Runs}
\tablehead{
\colhead{Run} &
\colhead{Dates} &
\colhead{Telescope} &
\colhead{Instrument} & \colhead{Number of Obs.}
}
\startdata
1 & 13-19 January 2016 & Gemini-N & DSSI & 6 \\
2 & 27-28 January 2016 & LDT & DSSI & 18 \\
3 &11 November 2016 & WIYN & NESSI & 1 \\
4 & 04-08 May 2017 & LDT & DSSI & 46 \\
5 & 09-11 June 2017 & Gemini-S & DSSI & 6 \\
6 & 18-21 October 2017 & LDT & DSSI & 16 \\
7 & 28 January-01 February 2018 & LDT & DSSI & 53 \\
8 & 27 August 2018 & LDT & DSSI & 21 \\
9 & 18-19, 21, 25 January 2019 & WIYN & NESSI & 134 \\
10 & 11-12 September 2019 & LDT & DSSI &77  \\
\enddata

\end{deluxetable}

\subsection{Observational Routine}

Observations were taken using the same techniques as detailed in earlier papers 
in this series. Specifically, objects in the target list were ordered in right ascension 
and then put into small groups at similar declinations. An unresolved bright
star was chosen as a point source calibration
object for each group from {\it The Bright Star Catalogue} \citep{hof82}. 

Once observations began on a given night, the telescope was kept close to the meridian to
minimize residual atmospheric dispersion, and objects were observed quickly,
usually with a cadence of approximately 2-10 minutes on each target, depending on its
brightness. Speckle data frames are taken at a standard exposure time of 40 ms at 
the LDT and at WIYN, and 60 ms at Gemini.
Data files are stored in 1000-frame stacks in FITS format. For fainter objects, three 
to five separate data
files were taken prior to moving on to the next object. Because both the DSSI and NESSI
speckle cameras take data in two different filters simultaneously, two such data files are
produced per 1000-frame sequence. 

\subsection{Data Reduction}

To reduce the data, the methodology remains the same as in Paper II of this series
\citep{hor11}, where the diffraction-limited Fourier transform of object intensity distribution
is assembled in the Fourier domain, low-pass filtered to reduce noise, and inverse transformed.
The modulus of this function is calculated separately from its phase. The former is derived by
computing the autocorrelation of each data frame, summing these, and Fourier transforming.
This results in the spatial-frequency power spectrum of the image frames. To arrive at the diffraction-limited
modulus, it is divided by the power spectrum of the point source calibration object and then the
square-root is taken. For the phase, we compute subplanes of the image bispectrum following
the standard method found in \citet{loh83}. Using these subplanes, 
we derive the phase from them using the relaxation technique of \citet{men90}.

Secondary stars for our systems are identified visually using the reconstructed images produced.
The pixel coordinate of the peak of the secondary is used as the starting point for the fitting of 
interference fringes in the power spectrum using a downhill simplex algorithm. This is described 
more fully in \citet{hor96}. 
The power spectrum fitting results in the position angle, separation, and
magnitude difference of the pair. 

\subsection{Pixel Scale and Orientation}

The pixel scale determination at all telescopes represented in Table 1 was carried out using 
a small group of calibration binary stars, that is, binaries with extremely well-known orbital elements,
usually determined with a preponderance of data from long baseline optical interferometry (LBOI)
observations. A sequential list of the objects, observation dates, and orbits used for this purpose is 
shown in Table 2. Generally, this
yielded results that were uncertain in terms of the scale at the $\sim$0.3\% level, and for the offset angle
between celestial coordinates and the pixel axes, the precision was approximately $\sim$0.2$^{\circ}$.

\begin{deluxetable}{rclrcrrrrl}
\tabletypesize{\scriptsize}
\tablewidth{0pt}
\tablenum{2}
\tablecaption{Orbits and Residuals Used in the Scale Determinations}
\tablehead{
\colhead{Run} &
\colhead{WDS} &
\colhead{Discoverer} &
\colhead{HIP} &
\colhead{Julian} &
\colhead{$\Delta \theta_{A}$} &
\colhead{$\Delta \rho_{A}$} &
\colhead{$\Delta \theta_{B}$} &
\colhead{$\Delta \rho_{B}$} &
\colhead{Orbit Reference} \\
&& \colhead{Designation} &&
\colhead{Year} &
\colhead{($^{\circ}$)} &
\colhead{(mas)} &
\colhead{($^{\circ}$)} &
\colhead{(mas)} &
}
\startdata
1 & $04136+0743$ & A  1938 & 19719 &
       2016.0324 & $+0.1$ & $-0.1$ & $+0.1$ & $-0.1$ & \citet{mut10} \\
      & &&&
       2016.0431 & $-0.1$ & $-0.4$ & $-0.1$ & $-0.3$ & \\
1 & $22409+1433$ & HO 296AB & 111974 &
       2016.0349 & $0.0$ & $-0.3$ & $-0.1$ & $-0.2$ & \citet{mut10} \\
    &   &&&
       2016.0431 & $0.0$ & $+0.9$ & $-0.1$ & $+0.7$ & \\
2 & $13100+1732$ & STF 1728AB & 64241 &
       2016.0704 & $+0.0$ & $+0.0$ & $+0.0$ & $+0.0$ & \citet{mut15} \\        
3 & $04136+0743$ & A  1938 & 19719 &
       2016.8567 & $+0.1$ & $-0.4$ & $+0.0$ & $+0.1$ & \citet{mut10} \\
3 & $21145+1000$ & STT 535AB & 104858 &
       2016.8560 & $-0.1$ & $+0.4$ & $+0.0$ & $-0.1$ & \citet{mut08} \\ 
4 & $13100+1732$ & STF 1728AB & 64241 &
       2017.3410 & $+0.2$ & $-2.4$ & $+0.1$ & $-1.4$ & \citet{mut15} \\ 
       &   &&&
       2017.3463 & $+0.2$ & $+3.0$ & $+0.2$ & $+2.5$ & \\ 
4 & $15232+3017$ & STF 1937AB & 75312 &
       2017.3413 & $+0.2$ & $-0.8$ & $+0.1$ & $-0.2$ & \citet{mut10} \\
4 & $15278+2906$ & JEF 1 & 75695 & 
       2017.3413 & $+0.0$ & $+0.1$ & $+0.0$ & $+0.1$ & \citet{mut10}  \\
4 & $17080+3556$ & HU 1176AB & 83838 &
       2017.3414 & $-0.6$ & $+0.2$ & $-0.4$ & $-0.9$ &\citet{mut10}  \\
5 & $18384-0312$ & A 88AB & 91394 &
       2017.4312 & $-0.6$ & $-3.2$ & $-0.6$ & $-2.8$ &  \citet{har13} \\
5 & $19026-2953$ & HDO 150AB & 93506 &
       2017.4395 & $+0.7$ & $+3.0$ & $+0.5$ & $+3.0$ & \citet{dro12} \\
6 & $21145+1000$ & STT 535AB & 104858 &
       2017.7951 & $0.0$ & $-0.1$ & $0.0$ & $-0.2$ & \citet{mut08} \\ 
6 & $22409+1433$ & HO 296AB & 111974 &
       2017.7952 & $0.0$ & $+0.1$ & $0.0$ & $+0.2$ & \citet{mut10} \\
7 & $04136+0743$ & A  1938 & 19719 &
       2018.0825 & $0.0$ & $0.0$ & $0.0$ & $0.0$ & \citet{mut10} \\
8 & $21145+1000$ & STT 535AB & 104858 &
       2018.6523 & $0.0$ & $0.0$ & $-0.1$ & $0.0$ & \citet{mut08} \\ 
9 & $13100+1732$ & STF 1728AB & 64241 &
        2019.0599 & $0.0$ & $-0.1$ & $0.0$ & $0.0$ &  \citet{mut15} \\ 
10 & $21145+1000$ & STT 535AB & 104858 &
       2019.6927 & $0.0$ & $+0.1$ & $0.0$ & $+0.1$ & \citet{mut08} \\ 
10 & $22409+1433$ & HO 296AB & 111974 &
       2019.6955 & $+0.1$ & $+0.1$ & $-0.1$ & $-0.5$ & \citet{mut10} \\
 \enddata
 
\end{deluxetable}




\section{Results}

Our final table of results obtained using the methods above is shown in Table 3.
The columns give:
(1) the Washington Double Star (WDS) number \citep{mas01}\footnote{{\tt http://astro.gsu.edu/wds/}}, 
which also
gives the right ascension and declination for the object in J2000.0 coordinates;
(2) an identifier from a standard star catalog, usually the Bright Star Catalogue ({\it i.e.,\ }Harvard Revised [HR]) number, 
the Henry Draper Catalogue (HD) number, or the Durchmusterung (DM) number of the object;
(3) the Discoverer Designation;
(4) the {\it Hipparcos} Catalogue number;
(5) the Julian year of the observation;
(6) the position angle ($\theta$) of the
secondary star relative to the primary, with North through East defining the
positive sense of $\theta$; (7) the separation of the two stars ($\rho$), in
arc seconds; (8) the magnitude difference ($\Delta m$)
of the pair in the filter used;
(9) the center wavelength of the filter ($\lambda_{c}$); and
(10) the full width at half maximum of the filter transmission ($\Delta \lambda$).
Note that, although observation epochs were stated in Besselian years in our previous 
papers in this series, here we have used Julian years, to align with IAU recommendations. 
If the Besselian date is needed, it can be obtained by using
\beq
BY = (JY - 0.041439661) \cdot 1.000021359.
\eeq

Thirty-one pairs in the table have no
previous detection of the companion
in the {\it Fourth Catalogue of
Interferometric Measures of Binary Stars} \citep{har01a}\footnote{{\tt http://astro.gsu.edu/wds/int4.html}}
and we therefore propose discoverer designations of
LSC (Lowell-Southern Connecticut) 131-161 here. (This continues the collection
of LSC discoveries detailed in Paper IX of this series.) All of these objects have
been confirmed as binary in at least one other observation not appearing here; 
a future paper will present the relative astrometry and photometry for those
confirming observations.

\begin{deluxetable}{lllrlrrrrl}
\tabletypesize{\scriptsize}
\tablewidth{0pt}
\tablenum{3}
\tablecaption{Binary star speckle measures}
\tablehead{
\colhead{WDS} &
\colhead{HR,ADS} &
\colhead{Discoverer} & HIP &
\colhead{JY} &
\colhead{$\theta$} & \colhead{$\rho$} &
\colhead{$\Delta m$} &
\colhead{$\lambda_{c}$} &
\colhead{$\Delta \lambda$} \\
\colhead{($\alpha$,$\delta$ J2000.0)} &
\colhead{DM,etc.} & \colhead{Designation} &&
\colhead{(2000+)} &
\colhead{($^{\circ}$)} &
\colhead{(${\prime \prime}$)}
 & \colhead{(mag)}
 & \colhead{(nm)} &
\colhead{(nm)}
}
\startdata
$00126+4419$ & HD 802   &  LSC 5 &  1011 &
19.0488  &  314.0 & 0.5792 & $<$4.64 & 562 & 44 \tablenotemark{a} \\
&&&&
19.0488 &   313.7 & 0.5821 & 4.90 & 832 & 40 \\
$00174+1852$ & BD+18 24 & HDS 39 & 1389 &
17.7953 &  247.4 & 1.5883 & $<$2.54 & 692 & 40 \tablenotemark{a} \\
&&&&
17.7953 &  247.5 & 1.5897 & $<$1.69 & 880 & 50 \tablenotemark{a}\\
$00182+5225$ & HD 1384   &  YSC 79   & 1460 &
19.0488  &   333.2 & 0.3112 & 3.45 & 562 & 44 \\
&&&&
19.0488 &   333.4 & 0.3140 & 3.38  & 832 & 40 \\
\enddata

\tablenotetext{a}{The magnitude difference appears as an upper limit because the observation
may be affected but the speckle decorrelation effect discussed in the text.}
\tablenotetext{b}{Quadrant determination inconsistent with other published measures.}
\tablenotetext{b?}{\hspace{3pt}Quadrant determination possibly inconsistent with other published measures.}
\tablecomments{Table 3 is published in its entirety in the machine-readable format.
      A portion is shown here for guidance regarding its form and content.}
\end{deluxetable}

Figure 1 visually summarizes the aggregate results from Table 3 in two ways. 
In Figure 1(a), we plot the magnitude
difference obtained as a function of the log of the separation. A curve that
illustrates our basic detection capability as determined in Paper IX \citep{hor20} is plotted with the data; 
this has two quasi-linear 
portions of different slopes and assumes a detection limit of $\Delta m = 0$ at the diffraction limit. 
The curve increases steeply from the diffraction
limit, but a ``knee'' is seen at approximately 0.1 arc seconds. 
For separations above this, the slope flattens out and
has is more modest throughout the remainder of the separation axis, until detection is no longer
possible due to the field-of-view restrictions on the speckle cameras we used. We see in the figure that
the current data set fills the presumed discovery space of DSSI subject to the conservative
assumption that the detection limit curve must go to a magnitude difference of zero at the diffraction limit.
In addition, some results appear at smaller separations than the diffraction limit with magnitude differences
higher than the curve shown. 
We have argued in past papers that, 
when systems are known to be binary through other observations,
the fits we obtain below the diffraction 
limit yield reasonable results, albeit with some decrease in precision, as in e.g.\ \citet{hor20}. 
The small-separation points above the blue detection limit curve drawn fit into that category. The black line in the
diagram continues the more modest slope to smaller separations, and the data indicate
that this may be a truer estimate of the sensitivity than the steeper slope, at least down to separations
as small as the diffraction limit.
In Figure 1(b), the magnitude difference obtained is shown as a function of the system $V$ magnitude;
we see here that virtually all systems reported here are brighter than 12th magnitude.
In these data, we do not detect a dependence in dynamic range sensitivity for V = 5--12.

\begin{figure}[!t]
\figurenum{1}
\plottwo{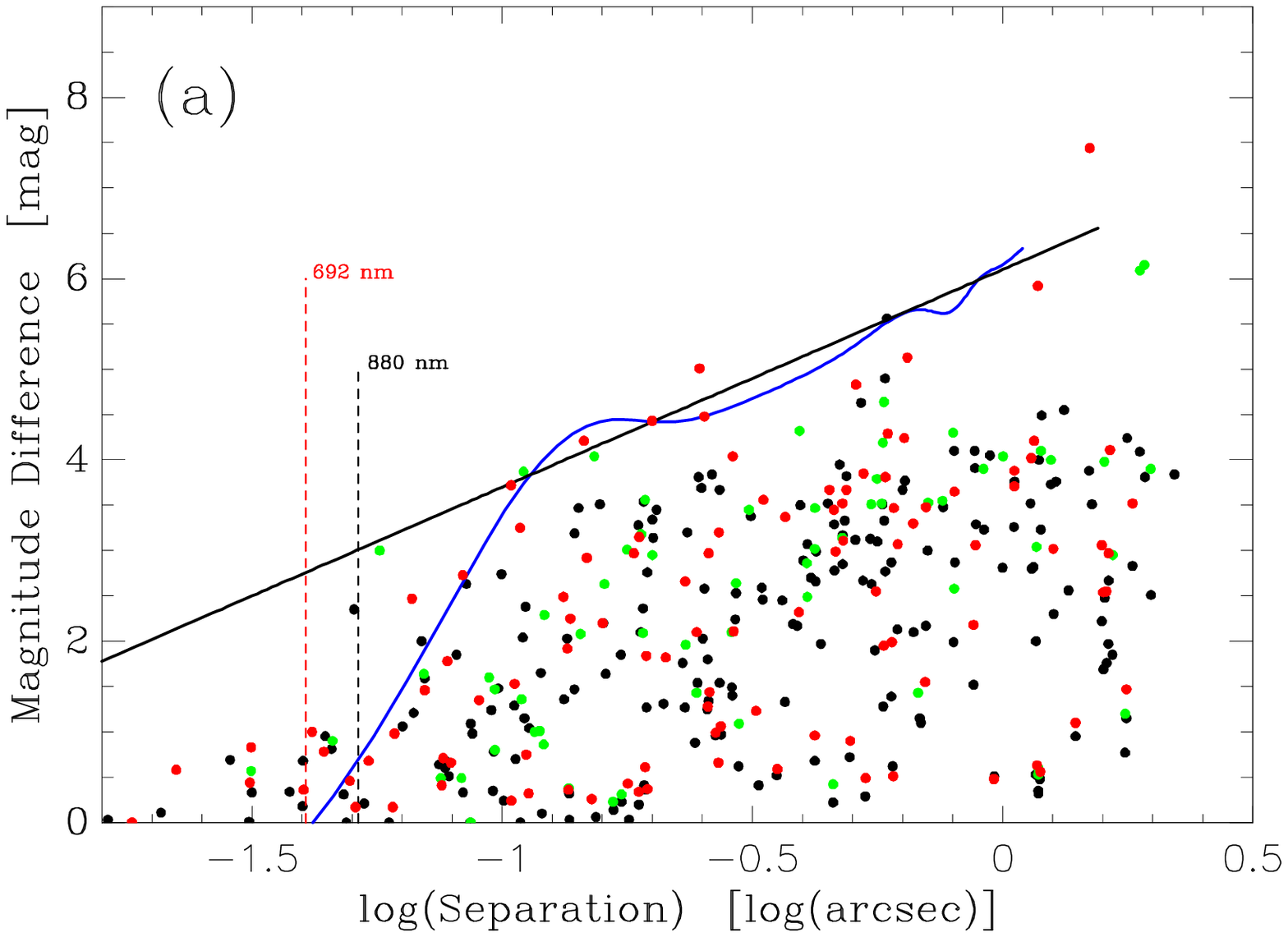}{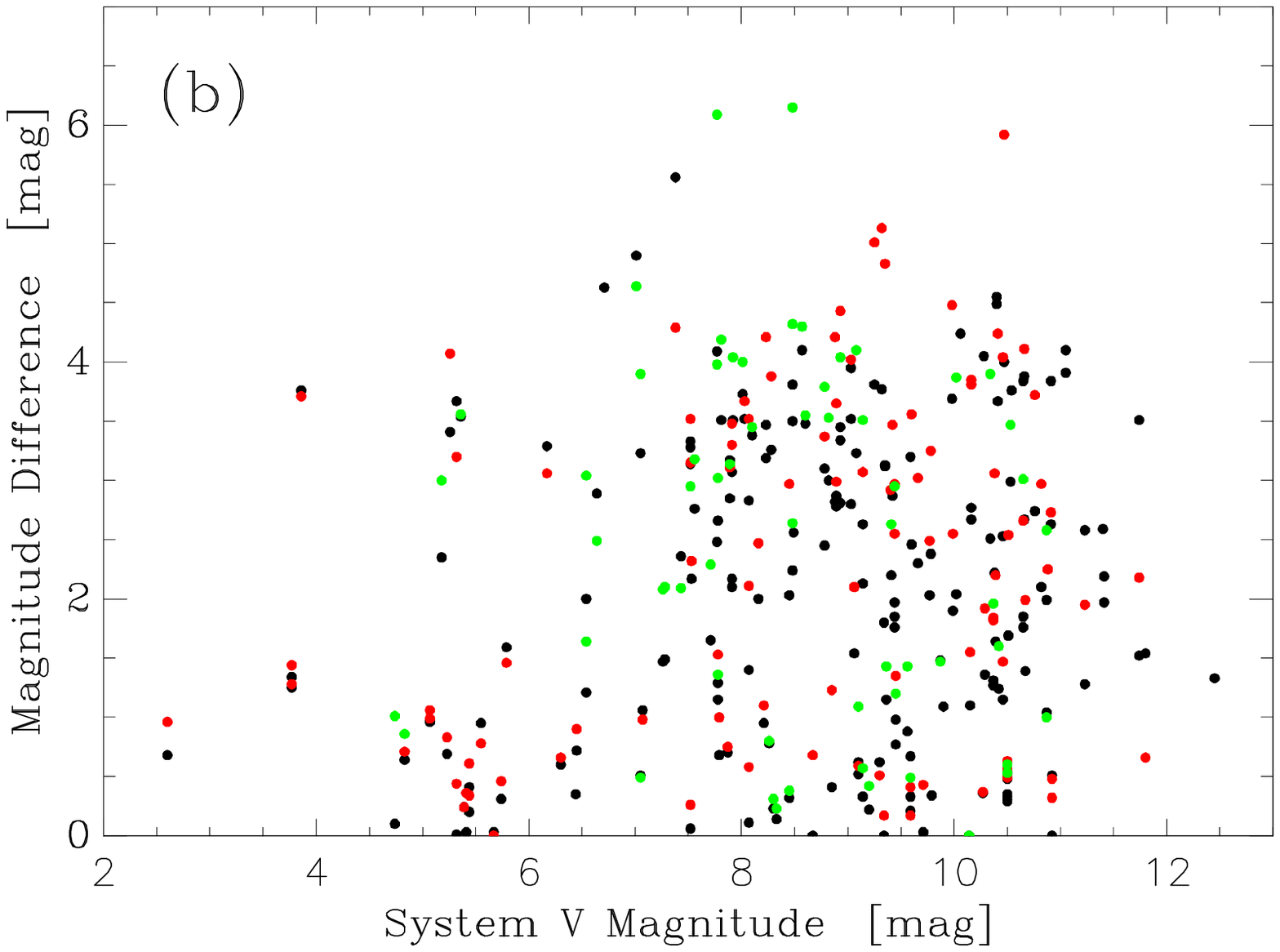}
\figcaption{(a) Magnitude difference as a function of separation for the observations in Table 3.
(b) Magnitude difference as a function of system $V$ magnitude. In both panels, the color
of the plot symbol indicates the filter wavelength used for the observation: green is 562 nm, 
red is 692 nm, and black is either 832 nm (for NESSI observations) or 880 nm (for DSSI
observations). In (a), the blue curve indicates a detection limit curve for LDT observations,
and the black line is drawn such that it matches the blue curve above a separation of 0.1 arcseconds.}
\end{figure}

\section{Analysis of the Data}

\subsection{Astrometric Precision}

\begin{figure}[!t]
\figurenum{2}
\plottwo{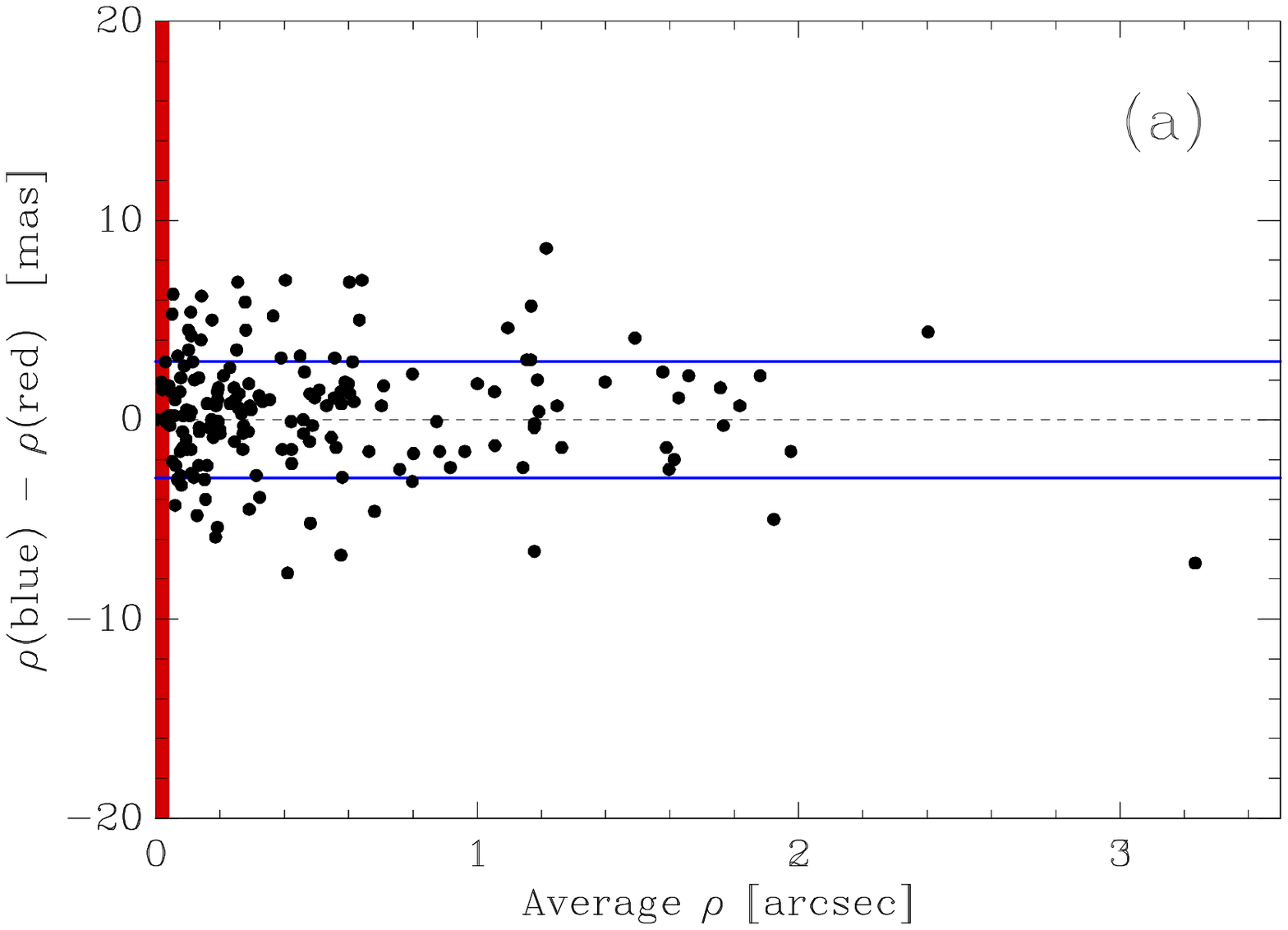}{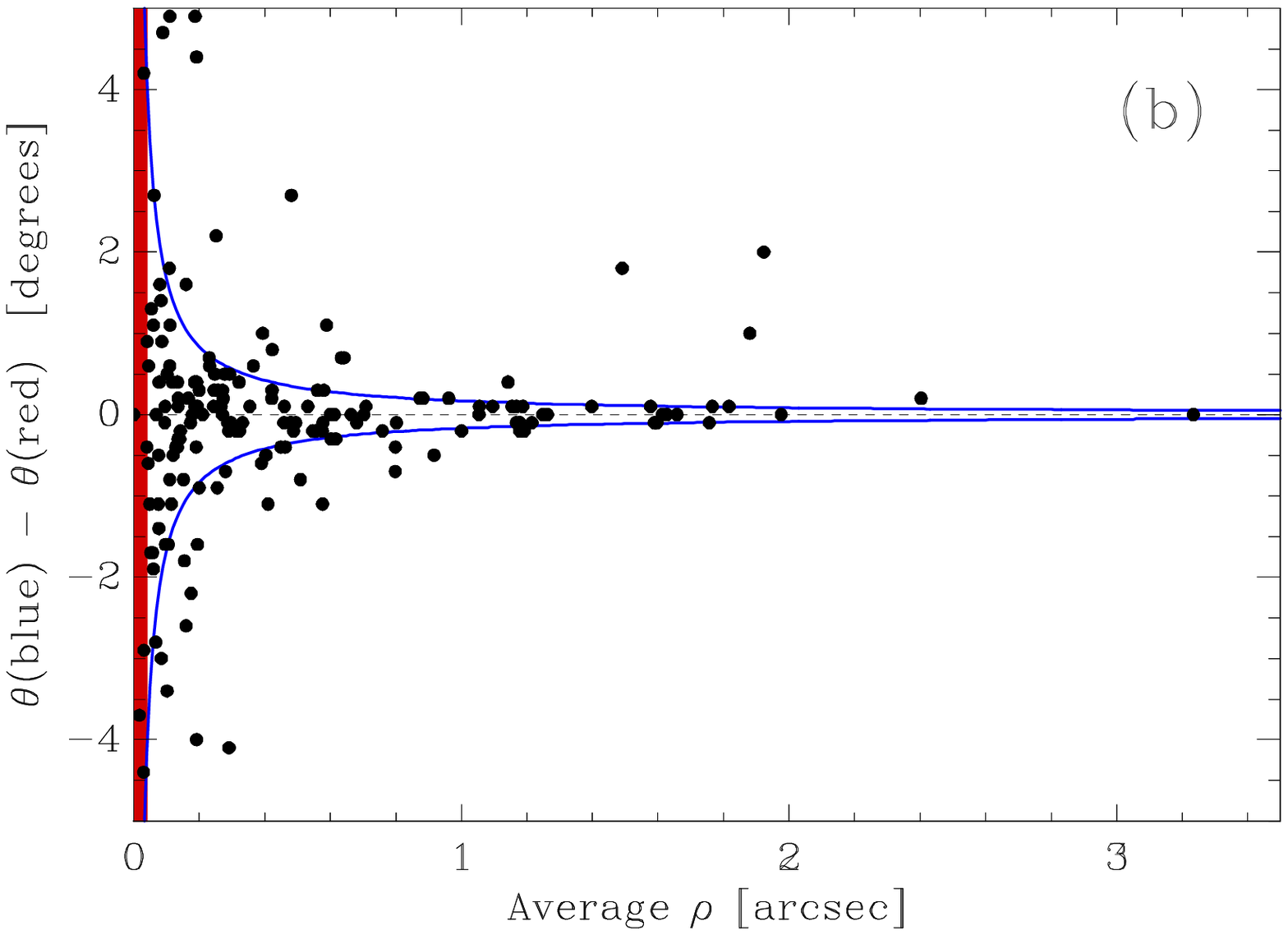}

\vspace{1cm}

\plottwo{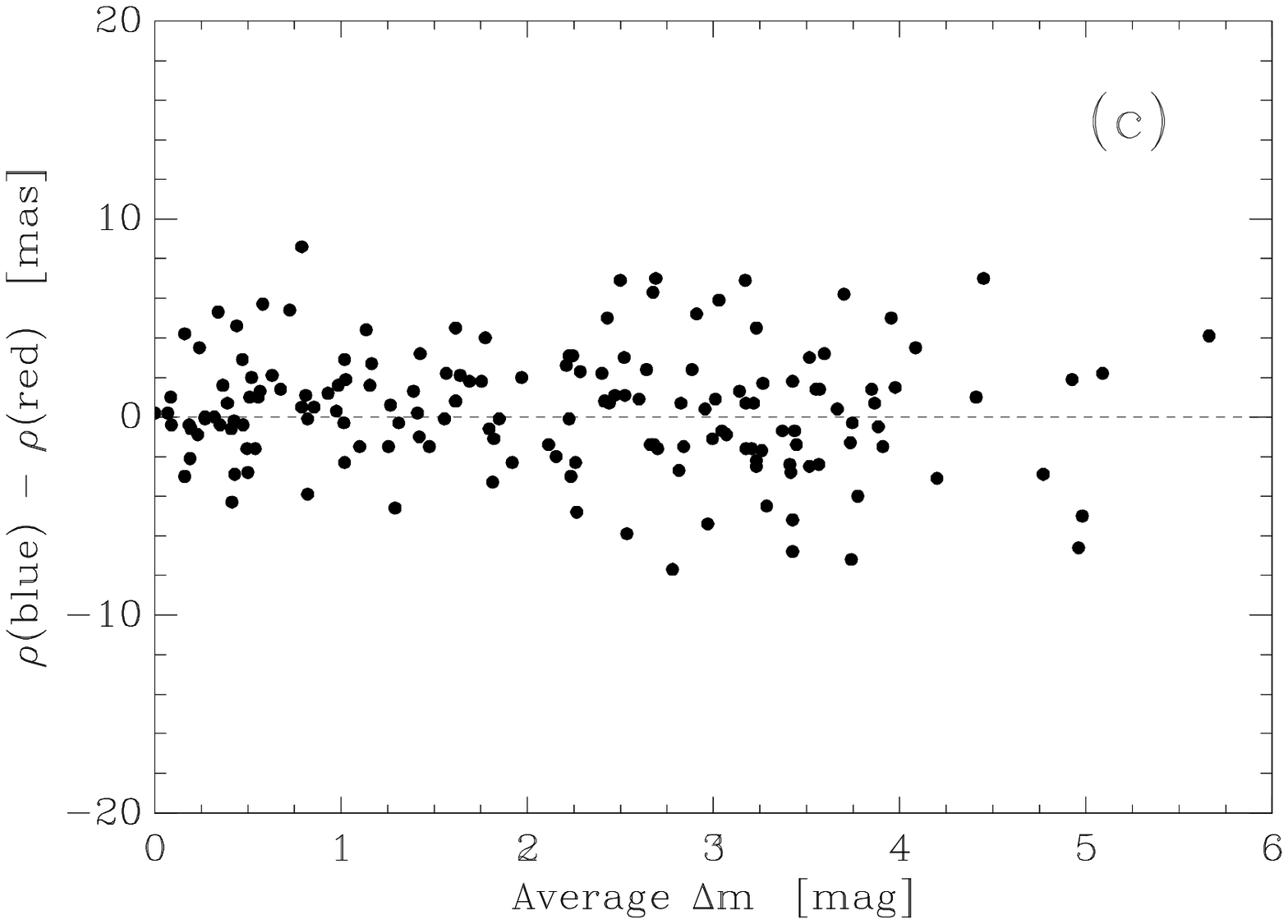}{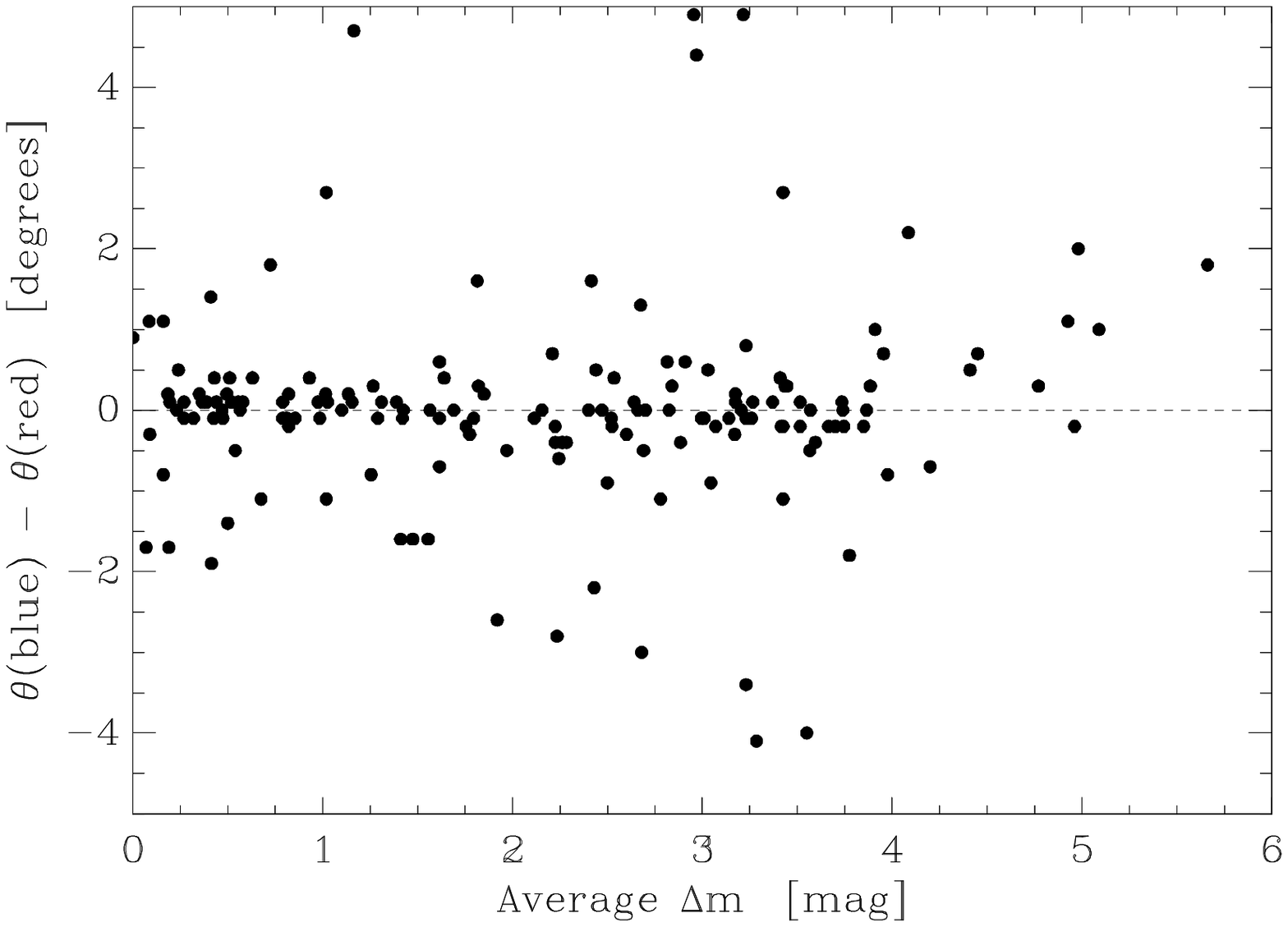}
\figcaption{Differences in the astrometric results obtained between paired observations at
the same epoch. (a) Differences in separation as a function of
average separation. (b) Differences in position angle as a function of average separation. 
In both plots, a dashed line at a difference of zero is drawn to guide the eye, and the blue
curves indicate the $\pm$1-$\sigma$ in estimated internal repeatability of individual measures
 as a function of separation as discussed in the text. For differences in separation, that is simply 
 the standard deviation of the measures, and for position angle, it is proportional to $\arctan{\delta \rho / \rho}$.
 In both (a) and (b), the red bar at the left marks the region below the formal diffraction limit.
 (c) Differences in separation as a function of magnitude difference. (d) Differences in position angle
 as a function of magnitude difference. In these latter two panels, a dashed line at zero is again drawn
 to guide the eye.}
\end{figure}

In order to characterize the astrometric precision of the measures in Table 3, we first utilize
the fact that, since DSSI and NESSI take simultaneous observations in two filters, most of
the measures listed in Table 3 are paired. Thus, we examine the
differences between the separation ($\rho$) and the position angle ($\theta$) obtained between
the two filters for the same epoch of observation. These are shown in Figure 2 for both coordinates
of the relative position. For separation, there is no obvious trend in either the differences 
or in their scatter as a function of average separation. These measures have an average difference of 
$0.40 \pm 0.22$ mas, and their standard deviation is $2.93 \pm 0.16$ mas. The former
perhaps indicates a slight systematic difference between the scale values
obtained in the two channels of the instrument. 
In contrast, the scatter in the difference in position angle
increases as the separation decreases; this indicates that at smaller separations, the same
positional uncertainty subtends a larger angle. 
The average value here is $-0.08 \pm 0.11^{\circ}$, indicating no
measurable offset between the two channels of the instrument. If we assume that
the positional uncertainty is the same in the orthogonal direction to the separation compared with the direction
of separation itself, we would expect the scatter in the position angle differences to vary as

\beq
\delta \theta = \arctan(\frac{\delta \rho}{\rho}) =  \arctan(\frac{2.93 {\rm mas}}{\rho}),
\eeq 

\noindent
where $\delta \theta$ is the uncertainty in the position angle difference and $\delta \rho$ is the 
uncertainty in
the separation difference. We have drawn this curve in Figure 2(b), and it appears to be consistent  with the 
scatter in the data. 
In Figure 2, we also plot the separation and position angle differences as a function of the average
of the magnitude difference obtained in the two channels of the instrument. These show a slight trend 
toward larger values as the magnitude difference increases, which is not unreasonable when the signal-to-noise
ratio of the secondary star decreases.
Since the data in all of the plots in Figure 2 are differences of two measures that presumably 
have similar uncertainties, we
can infer that the uncertainty of a given single measure in Table 3 is reduced from the above standard
deviation by a factor of $\sqrt{2}$, to $2.07 \pm 0.11$ mas, and the position angle uncertainty is similarly
reduced.

We also studied cases in Table 3 where an orbit determination exists in the literature and we compare
our measures with the ephemeris positions derived from the published orbital elements. We confine ourselves
to objects with orbits in the Sixth Orbit Catalog \citep{har01b}\footnote{\tt http://astro.gsu.edu/wds/orb6.html} 
of Grade 2 or better and which have uncertainties in the published
orbital elements available in the Sixth Orbit Catalog. We further restrict ourselves to those objects which
have observations in two filters at the same epoch in Table 3. Figure 3 shows the residuals obtained,
where we have averaged the position angle and separation between the channels for each pair of
observations and plot that as a single point in Figure 3. We have also drawn the same curves as 
in Figures 2(a) and (b), although since we have averaged two independent measures to obtain each 
data point here, we also divided the internal precision number by $\sqrt{2}$, which is how the resulting 
uncertainty would decrease for two truly independent samples of the same uncertainty. Thus,
the lines drawn in Figure 2 for separation were at $\pm$2.07 mas, here they are drawn at 
$\pm$2.07/$\sqrt{2}$ = $\pm$1.46 mas, and the curves drawn for the position angle are likewise modified.

\begin{figure}[!t]
\figurenum{3}
\plottwo{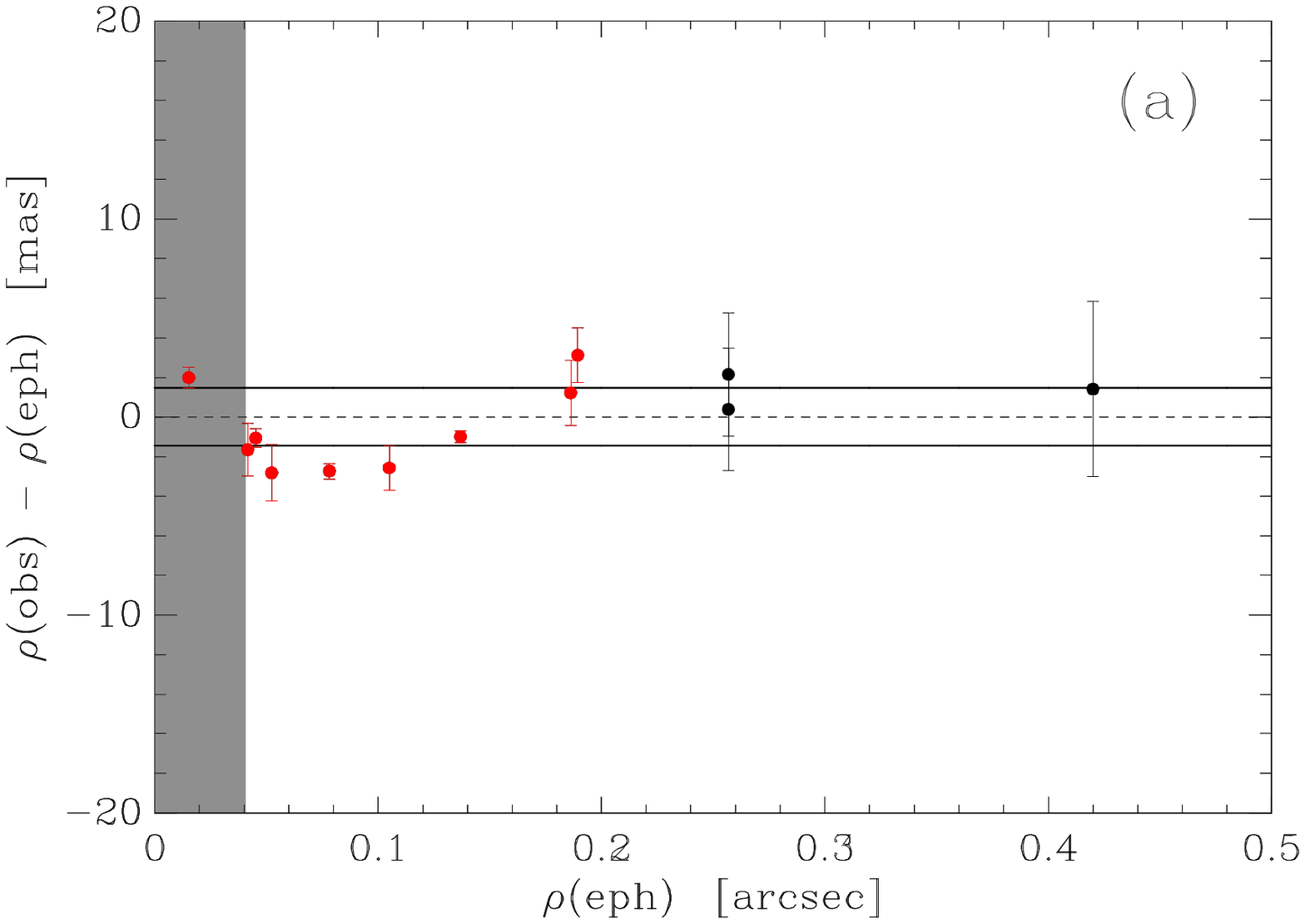}{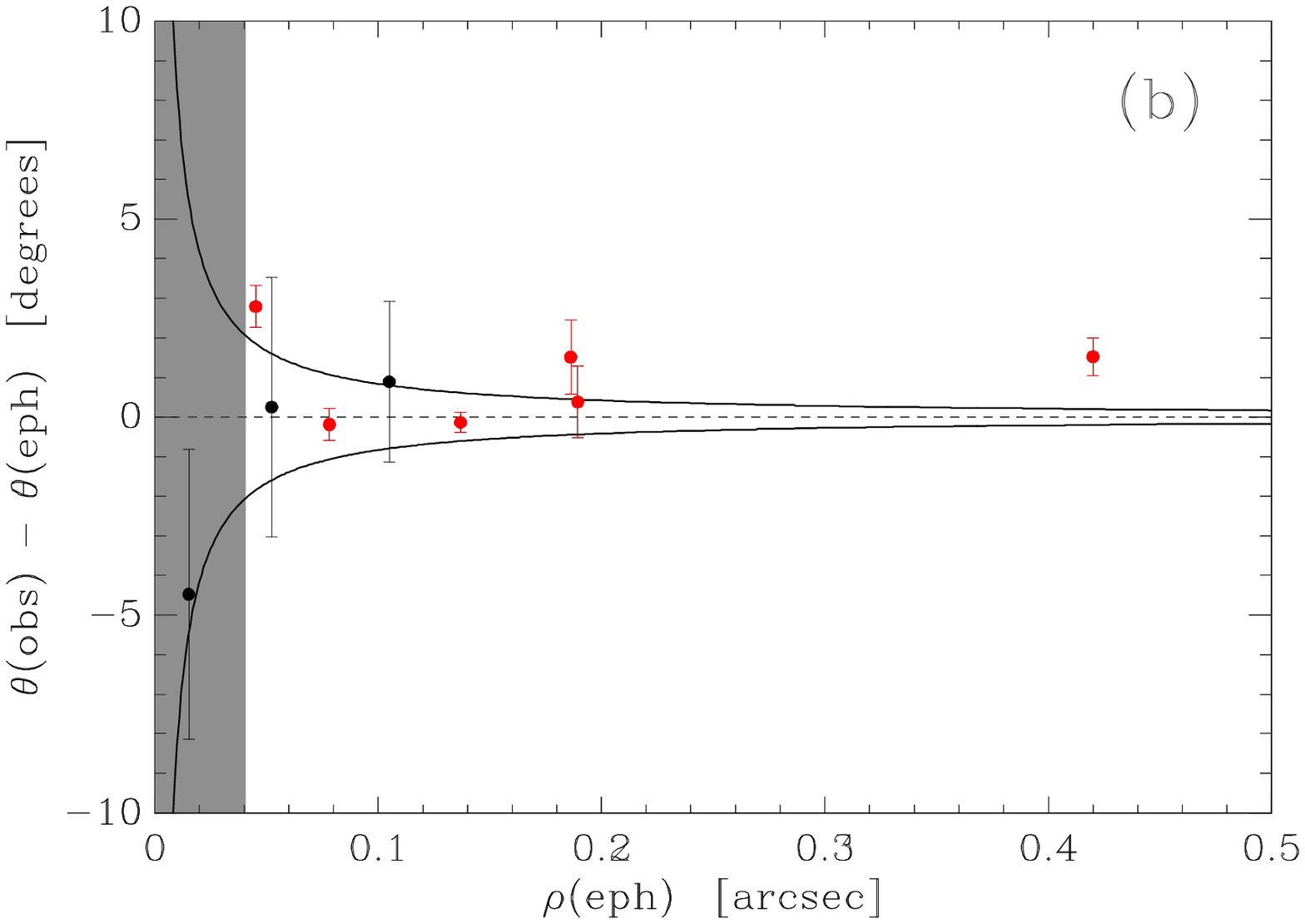}
\figcaption{Residuals obtained when comparing with ephemeris position for objects in 
Table 3 that have a Grade 1 or Grade 2 orbit in the Sixth Orbit Catalog. The error bars drawn are those of the 
ephemeris prediction, based on the published uncertainties in the orbital elements. (a) Separation
residuals as a function of average separation.  The red points are objects with
ephemeris uncertainties of less than 2 mas.
 (b) Position angle
residuals as a function of average separation. Here, 
the red points indicate position angle uncertainties of less than 1$^{\circ}$.
In both plots, the expected 1-$\sigma$ uncertainty from our internal precision study is 
drawn as discussed in the text, and the grey regions at the left of each plot 
mark separations below the formal diffraction limit of the LDT at 692 nm.}
\end{figure}

The results indicate complete consistency with the
internal repeatability study. 
For separation, the average is $-0.13 \pm 0.61$
mas with a standard deviation of $2.10 \pm 0.43$ mas. 
In position angle, we find an average residual of $0.06 \pm 0.51^{\circ}$
and a standard deviation of $1.78 \pm 0.36^{\circ}$. We conclude that there are no identifiable
sources of systematic error in the data set, and that our individual measures in Table 3
have average uncertainty of 2.07 mas in separation and $\arctan(2.07 {\rm mas}/\rho)$ in position angle.


\subsection{Photometric Precision}

Regarding the photometric precision of our measures, we have compared our results to space-based measures
in two ways. First, as in previous papers in this series, we use data in the {\it Hipparcos} Catalogue,
although the $H_{p}$ filter is not a good match for any of the filters we use in speckle imaging as it
is wider and bluer than any of the filters we have used in our observations. In addition, a number of
the stars in Table 3 do not have component information in the {\it Hipparcos} Catalogue. On the other hand,
high-precision data are available from {\it Gaia} in the $G$ and $R_{p}$ filters,
which are reasonably similar to our 562- and 692-nm filters, respectively. 
However, for separations below 1 arc
second, which is the vast majority of our sample, there is almost nothing available in EDR3. Both 
samples are further complicated by the presence of triples, which we resolve in our observations
but are not resolved in the {\it Gaia} or {\it Hipparcos} results. Examples in this category not
used for a comparison. We also removed one other object, HIP 63942,
because the uncertainty in the flux in the {\it Gaia} $R_{p}$ filter was very large in EDR3, and we do
not consider systems with $\Delta m$ less than 0.25, as the speckle magnitude differences typically have
larger scatter and uncertainties about quadrant determinations in that range. Setting all of these objects aside, 
we are left with a relatively small sample of only 33 observations to study in detail. 

Nonetheless, we show a comparison between the space-based and speckle results in Figure 4.  
In panel (a), we plot the difference between the speckle $\Delta m$ and the space-based value as a
function of seeing times separation of the speckle observation. In previous papers, we have used this
quantity as a way to judge the isoplanicity of the observations, and thus the reliability of the photometry.
The larger that this quantity is, then the less the speckle patterns between the primary and secondary stars
will resemble each other, so that in the standard analysis, a secondary will appear fainter than
it actually is. Therefore, the difference between the speckle $\Delta m$ and that obtained from a space-based
source should be near zero for small values of seeing times separation, and grow as this quantity
gets larger. This is indeed the general trend in Figure 4(a). 

As in our previous papers, if we confine our attention to those observations in Table 3 that have seeing
times separation less than 0.6 arcsec$^{2}$, then these should be relatively unaffected by non-isoplanicity, and
therefore correlate well with the space based values. This is shown in Figure 4(b). It may be seen here
that all of the objects in this subsample are {\it Hipparcos} stars, and some have sizable uncertainties.
However, within the uncertainty, the correlation is quite good. The average uncertainty $\delta(\Delta
H_{p})$ for the points in Figure 4(b) is 0.327 magnitudes; if we subtract this value in quadrature from the 
standard deviation of the plot residuals, an estimate of the typical uncertainty in the speckle $\Delta m$
can be made; this results in a value of 0.092 magnitudes, which is comparable to what we have quoted
in earlier papers in this series (typically 0.1 to 0.15 magnitudes). We conclude that, though this result is
not as statistically robust as in the larger samples studied in previous papers, it remains consistent with
them in terms of the photometric precision of the measures we present. Although we did not compare 
our results at 832 and 880 nm directly with the space-based measures, there are two cases in Table 3
where we report three individual measures at these wavelengths (HIP 55605 and HIP 57058). 
In these cases, the average standard deviation in the delta-m is $0.13 \pm 0.09$, indicating that at least the 
internal precision appears comparable to the other wavelengths, to the extent that we can determine.

\begin{figure}[!t]
\figurenum{4}
\plottwo{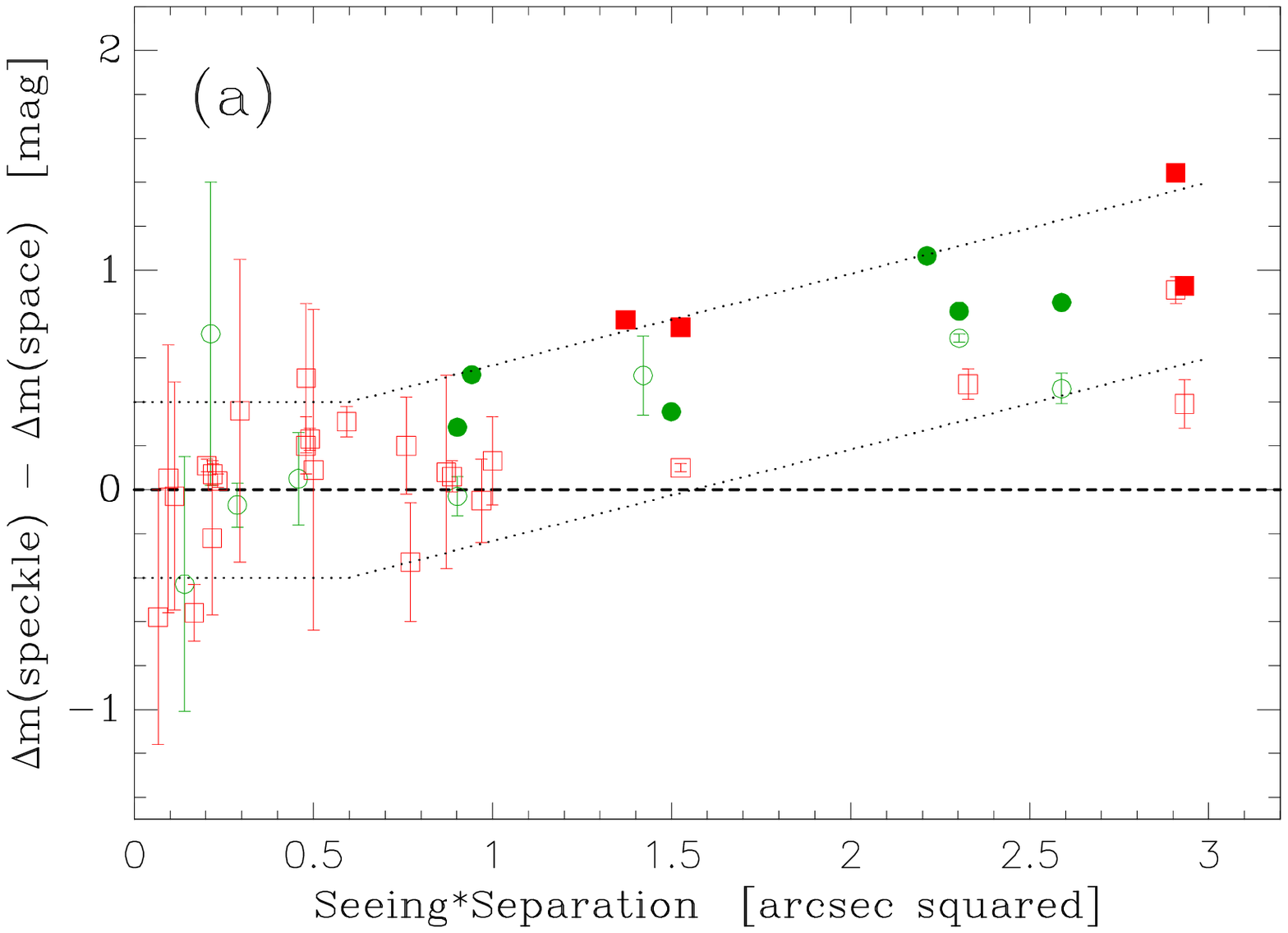}{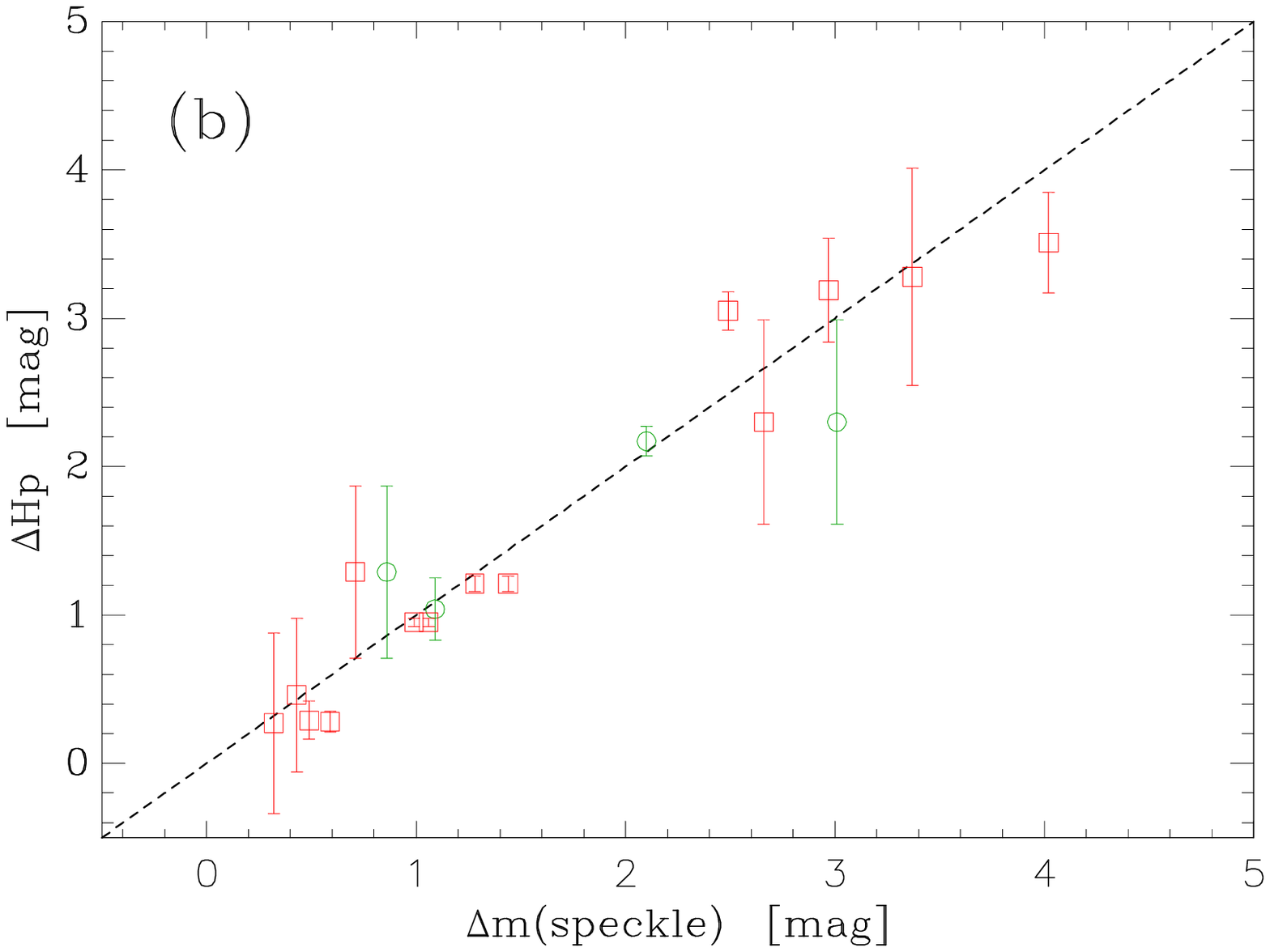}
\figcaption{A comparison of speckle magnitude difference versus the {\it Gaia} or {\it Hipparcos} magnitude
difference. (a) The difference of these two measures as a function of seeing times separation of the
speckle observation. (b) For those objects with seeing times separation below 0.6 arcsec$^{2}$ the
space-based magnitude difference is plotted as a function of the speckle magnitude difference. In 
both plots, red square symbols indicate that the speckle magnitude difference is in the 692-nm filter, 
and green circles indicate that the speckle magnitude difference is in the 562-nm filter. The symbols
are drawn as filled in either case if the uncertainty in the space-based measure is less than 0.1 magnitudes.}
\end{figure}

\section{Orbit Calculations}

A set of 25 systems in Table 3 have shown significant relative motion since they were discovered, 
and the new data, together with previous measures in the literature, permit the calculation of first orbital 
elements in some cases. We use the orbit code described in \citet{mac04} for this purpose,
which performs a grid search based on a low and high value of each orbital element provided, and
then after determining the best fit to the data on the grid, it performs a downhill simplex calculation
to reach the global minimum in reduced-$\chi^2$. To estimate uncertainties in the final orbital elements, 
we perform the orbit calculation many times, typically 50 to 100 times, and in each case we throw in
normally-distributed offsets to the position angle and separation values used in the calculation.
The standard deviations of these offsets is 2.5 mas in separation and 0.0025/$\rho$ degrees in
position angle. Of course, the uncertainties obtained should still be viewed as formal and not necessarily
representing the true uncertainty of each orbital element; most of the orbits presented are fits 
to only a portion of the orbital path or represent sparse coverage of the orbital ellipse. Further 
observations will be needed to refine these orbits, and we highlight them here in an attempt to 
encourage other observers to take more data on these systems in the coming years.

\begin{deluxetable}{llrlrrrrrrr}
\tabletypesize{\scriptsize}
\tablewidth{0pt}
\tablenum{4}
\tablecaption{Visual Orbital Elements for 25 Systems}
\tablehead{
\colhead{WDS} &
\colhead{Discoverer} &
\colhead{HIP} &
\colhead{Spectral} &
\colhead{$P$} &
\colhead{$a$} &
\colhead{$i$} &
\colhead{$\Omega$} &
\colhead{$T_{0}$} &
\colhead{$e$} &
\colhead{$\omega$} \\
& \colhead{Designation} && \colhead{Type\tablenotemark{a}} & \colhead{(yr)} &
\colhead{($^{\prime \prime}$)} &
\colhead{($^{\circ}$)} &
\colhead{($^{\circ}$)} &
\colhead{(JY)} &
\colhead{} &
\colhead{($^{\circ}$)}}

\startdata
$00132+2023$ &HDS  29  & 1055 & K7V &  78.8   & 0.4233  & 136.4   & 331.7  & 2062.4    & 0.895  & 348.3  \\
 &        &&        &   2.3   & 0.0067  &   7.6   &   3.1  &    2.1    & 0.021  &   4.0  \\
$02164+0438$ & YR     8   & 10596 & F0V &   48.9  &   0.1080  &  101.7  &   332.94 &  2055.   &    0.324  &   82. \\
  &        &&        &     6.7   &  0.0095    &  1.9    &   0.89  &  21.    &    0.077   &  24. \\
$02167+0632$ &YSC   20 & 10616 & K0 &  13.61  & 0.1255  &  22.9   & 192.   & 2010.705  & 0.4529 & 248.   \\
   &      &&        &   0.14  & 0.0017  &   4.0   &  12.   &    0.032  & 0.0078 &  12.   \\
$03376+2121$ &PAT    1   & 16908  & K0V & 110.0   &  0.8475  &  111.61  &  292.0  &  2071.06  &   0.434  &  251.66 \\
  &        &&        &     1.4  &   0.0081   &   0.32   &   1.1   &    0.60   &  0.016   &   0.71 \\
$04268+1240$ & WOR   15  &  20745  & K2    & 40.9  &   0.314   &  144.  &    136.   &  2037.1  &    0.811  &  319. \\
  &         &&         &    3.3   &  0.047   &   14.    &   19.    &    3.3    &  0.024  &   16. \\
$05009+6107$ & HDS  650 & 23317 & K5 &   39.3   & 0.242   &  93.8   & 341.0  & 2058.3    & 0.49   & 196.   \\
  &       &&        &   4.7   & 0.025   &   1.5   &   2.0  &    8.6    & 0.16   &  27.   \\
$08289-1552$ & RST 4403 & 41609 &K2V & 205.6   & 1.191   & 160.0   &  67.5  & 2149.6    & 0.183  &  25.8  \\
 &        &&        &   2.4   & 0.021   &   5.9   &   2.2  &    3.8    & 0.034  &   2.8  \\
$08447-2126$ & HDS 1260A-BC & 42910 & K7V &  106.1  & 0.764   & 149.    & 248.7  & 2126.4    & 0.167  &  97.   \\
  &       &&        &    5.3  & 0.027   &  10.    &   8.8  &    3.1    & 0.066  &  13.   \\
$08447-2126$ & TOK 395BC & 42910 & M3V? & 9.06    & 0.1158  & 105.9   &   7.7  & 2008.2    & 0.551  & 350.   \\
  &      &&        & 0.58    & 0.0045  &   2.7   &   2.9  &    2.5    & 0.080  &  11.   \\
$10320+0831$ & YSC   39 &  51571 &K5V &  178.9   & 1.083   & 129.5   &  50.95 & 2000.07   & 0.274  & 331.4  \\
  &       &&        &   5.4   & 0.015   &   1.6   &   0.79 &    0.60   & 0.012  &   2.0  \\
$11114+4150$ & HDS 1593 & 54663 & K2 &    36.8   & 0.265   &  70.9   & 329.0  & 2000.2    & 0.269  &  74.6  \\
  &       &&        &   3.0   & 0.013   &   3.2   &   1.5  &    1.1    & 0.040  &   5.8  \\
$11235+0701$ & BAG  24Aa,Ab & 55605  & K4V &  20.80  &  0.227   &  160.   &   316.   &  2014.29   &  0.290  &  386. \\
   &       &&         &    0.11  &  0.010   &   12.    &   19.    &    0.57  &   0.030  &   22. \\
$11418+0508$ & LSC 141  &  57058   & K4V &   3.728 &  0.0807   &  32.6   &  110.2  &  2017.949  &  0.338  &  269.4 \\
   &        &&         &   0.057  & 0.0016   &   2.2    &   9.4    &   0.020   & 0.019   &   9.6 \\
$11471-1149$ & RST 3756 &  57494 & K4.5V &  128.0   & 1.164   & 132.15  & 151.6  & 2044.1    & 0.108  &  89.7  \\
  &       &&        &   2.7   & 0.010   &   0.96  &   2.0  &    1.5    & 0.027  &   5.5 \\
$11539+1402$ & YSC   96 &  58006 & F0 &   36.8   & 0.123   &  36.0   &  86.   & 2006.5    & 0.13   &  94.   \\
 &        &&        &   4.9   & 0.014   &   6.8   &  13.   &    9.9    & 0.13   &  25.   \\
$13331+4316$ & COU 1754 &  66110 & K8V &  125.5   & 0.6524  &  94.865 & 171.25 & 2056.4    & 0.189  & 230.13 \\
  &       &&        &   2.0   & 0.0059  &   0.073 &   0.34 &    1.2    & 0.018  &   0.21 \\
$13450+0206$ & HDS 1935 &  67086 & K5 & 209.    & 0.730   & 146.7   &   7.7  & 2188.     & 0.270  & 209.   \\
  &       &&        &  18.    & 0.036   &   4.0   &   9.7  &   21.     & 0.064  &  11.   \\
$14136+5522$ & HDS 1995 & 69488 & K0 &    40.9   & 0.333   & 127.3   & 241.6  & 2028.9    & 0.387  & 248.2  \\
  &       &&        &   1.7   & 0.013   &   1.8   &   4.2  &    1.6    & 0.047  &   3.1  \\
$14330+0656$ & YSC    6  &  71142  & K0 &    57.6  &   0.199   &   40.5   &  246.2 &  2044.  &     0.122  &   33. \\
  &        &&        &     9.0   &  0.025    &   6.0    &   3.8    &  12.    &   0.059   &  29. \\
$14136+5522$ & HDS 2211 &  76768 &  K5V &  79.5   & 0.592   & 136.6   & 272.8  & 2042.6    & 0.679  &  49.3  \\
 &        &&        &   2.3   & 0.015   &   3.2   &   3.5  &    1.1    & 0.023  &   4.0  \\
$17577-2143$ & HDS 2530   & 87925  & K6+V  &  54.9  &   0.5214   &  63.8   &  148.2  &  1999.5   &   0.585   & 249.82 \\
  &        &&        &     1.6   &  0.0073    &   1.5    &  1.4    &   0.5   &   0.013   &   0.75 \\
$19153+2454$ & HDS 2724  & 94622 & K5 &    8.450 & 0.152   & 116.4   & 284.9  & 2005.2    & 0.329  & 212.   \\
  &       &&        &   0.025 & 0.015   &   3.9   &   6.7  &    1.1    & 0.068  &  40.   \\
$19233-0635$ & HDS 2745 & 95299 & K3/4(V) & 161.7   & 0.893   & 145.9   &  18.1  & 2124.63   & 0.549  & 209.7  \\
  &       &&        &   1.0   & 0.016   &   3.6   &   1.0  &    0.59   & 0.031  &   1.8  \\
$19467+4421$ & YSC  136 & 97321 & F9IV-V &  4.813  & 0.0831  &  64.3   &  89.9  & 2010.524  & 0.180  & 261.0  \\
   &      &&        &  0.019  & 0.0022  &   1.3   &   1.7  &    0.047  & 0.022  &   3.7  \\ 
$23464-2302$ & TOK  375 & 117247 & K2+V &  12.5    & 0.1426  & 165.4   & 232.2  & 2022.47   & 0.429  & 149.2  \\
  &       &&        &  1.0    & 0.0037  &   5.9   &   8.0  &    0.78   & 0.044  &   9.0  \\
\enddata

\tablenotetext{a}{From SIMBAD \citep{wen00}.}
\end{deluxetable}

The visual orbital elements obtained are shown in Table 4, where
the orbits span a range in periods of 3--206 years and semimajor axes of 0.08--1.2 arcsec.
All of these are cases for
which the orbit has been calculated for the first time, and most stars have been previously 
identified as K dwarfs. The first column gives the discoverer designation from Table 3 followed by the
spectral type as it appears in SIMBAD\footnote{\tt https://simbad.u-strasbg.fr/simbad}
\hspace{0.05cm} \citep{wen00}. Column 3 shows
the {\it Hipparcos} number, and the remaining columns contain the seven 
standard visual orbital elements and their estimated uncertainties.

\subsection{The Triple System HIP 42910}

HIP 42910 was first discovered to be binary in observations taken with the {\it Hipparcos} satellite \citep{esa97},
yet no speckle measures of the pair were taken until our work at the WIYN telescope in 2012 \citep{hor17}.
That observation also revealed that the fainter component was a small-separation pair, but the work was 
published after observations of  \citet{tok15}, thus the BC component bears the discoverer designation of
those authors. Since 2012, a very nice sequence of speckle observations has been taken and 
published by both groups, including the most recent ones that are included in this paper in Table 3. The longer time 
baseline of the wider component (A-BC) allows for a reasonable orbit calculation there (with $P=106.1$ years), while the BC pair 
has a much shorter period and smaller separation, and therefore the current group of speckle
measures for that pair already covers about one third of the orbital ellipse ($P=9.06$ years). 
The orbits for both components are shown in Figure 5. 
SIMBAD
lists the composite spectral type as 
K7V and observations by the RAVE project \citep{kor13} indicate a 
metallicity near solar ([m/H] = $0.08 \pm 0.09$). 

\begin{figure}[!t]
\figurenum{5}
\plottwo{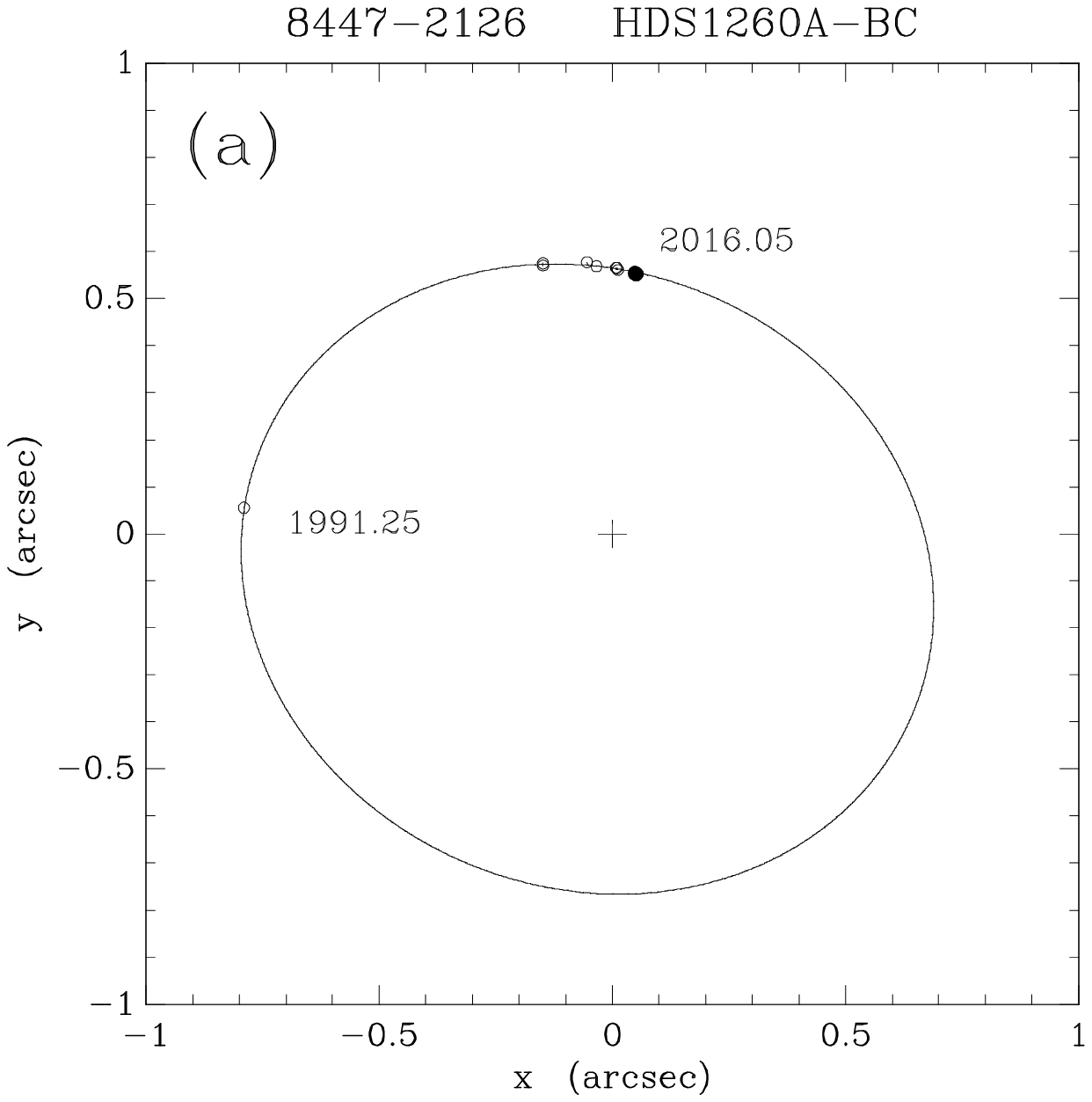}{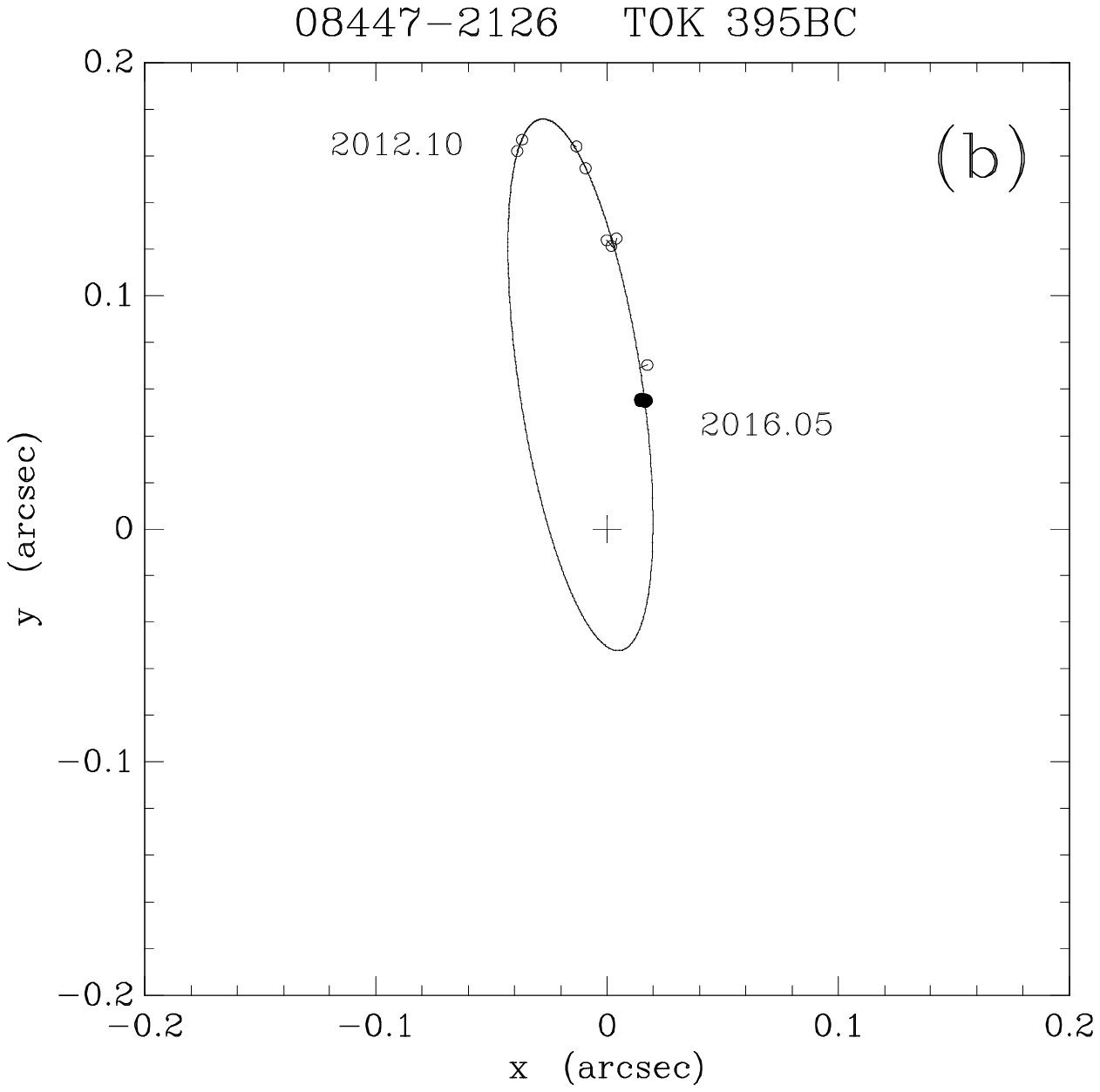}
\figcaption{Visual representations of the orbits in Table 4 for the triple system HIP 42910. 
(a) HDS 1260A-BC; (b) TOK 395BC. In all plots, north is down and east is to the right. 
The cross marks the origin in each panel. Measures appearing in Table 3 are shown
as filled circles while previous measures listed in the 4th Interferometric Catalog are drawn as open
circles, and line segments are drawn from ephemeris position on the orbit to each data point. The
positions shown for the BC component in panel (a) are obtained by averaging the published positions
of AB and AC for each epoch.}
\end{figure}

\subsection{A Spectroscopic-Visual Orbit for HD 173093 = YSC 133}

\subsubsection{Spectroscopic Observations}

Spectroscopic observations of HD~173093 (= HIP 91880 = YSC 133)
at Kitt Peak National 
Observatory (KPNO) began a decade before it was learned that this 
system was also being observed with speckle interferometry. 
Discovery of our mutual interest and our complementary observing
techniques led to the current effort to determine the orbital and physical
properties of the system.

From 2001 June through 2014 July we obtained 33 spectroscopic 
observations of HD 173093 at KPNO with the 0.9~m coud\'{e} feed 
telescope and spectrograph. The initial 11 observations were
acquired when the star was included in a solar-type star survey 
of \citet{aw06}. However, HD~173093 was dropped from that survey 
after the sample was limited to stars within 25 pc of the Sun. 
From those initial observations it was determined to be a 
double-lined binary, and so additional observations were 
obtained, which revealed the star to be a triple system.  

The KPNO spectra were acquired with various gratings and CCD 
detectors and so resulted in spectra that had different wavelength
ranges and different resolutions. 
The various combinations are listed in Table~\ref{tbl-instcomb}. 
Additional information is provided in \citet{aw06},
\citet{fetal19}, and \citet{hor20}.

\begin{deluxetable}{ccllc}
\tabletypesize{\footnotesize}
\tablewidth{0pt}
\tablecolumns{5}
\tablenum{5}
\tablecaption{Telescopes and Instruments for Spectroscopic Observations\tablenotemark{a}\label{tbl-instcomb}}
\tablehead{\colhead{Date} & \colhead{Helio. Julian Date} & \colhead{Telescope} &
\colhead{Instrument,} & \colhead{Resolution} \\
\colhead{d/m/yr} & \colhead{$-2400000$} & \colhead{} & \colhead{grating,CCD} &
\colhead{$\lambda/\Delta \lambda$ (2 pix.)}}
\startdata
13 06 2001 & 52073 & KPNO coud\'{e} feed & coud\'{e} spec.,A,F3KB & 31,250 \\
14 06 2001 & 52074 & KPNO coud\'{e} feed & coud\'{e} spec.,A,F3KB & 31,250 \\
15 06 2001 & 52075 & KPNO coud\'{e} feed & coud\'{e} spec.,A,F3KB & 31,250 \\
16 06 2001 & 52076 & KPNO coud\'{e} feed & coud\'{e} spec.,A,F3KB & 31,250 \\
08 05 2002 & 52402 & KPNO coud\'{e} feed & coud\'{e} spec.,A,F3KB & 31,250 \\
09 05 2002 & 52403 & KPNO coud\'{e} feed & coud\'{e} spec.,A,F3KB & 31,250 \\
28 09 2002 & 52545 & KPNO coud\'{e} feed & coud\'{e} spec.,A,F3KB & 31,250 \\
30 09 2002 & 52547 & KPNO coud\'{e} feed & coud\'{e} spec.,A,F3KB & 31,250 \\
09 10 2002 & 52556 & KPNO coud\'{e} feed & coud\'{e} spec.,A,F3KB & 31,250 \\
10 10 2002 & 52557 & KPNO coud\'{e} feed & coud\'{e} spec.,A,F3KB & 31,250 \\
11 10 2002 & 52558 & KPNO coud\'{e} feed & coud\'{e} spec.,A,F3KB & 31,250 \\
11 03 2003 & 52710 & KPNO coud\'{e} feed & coud\'{e} spec.,A,TI5 & 30,000 \\
16 06 2004 & 53172 & KPNO coud\'{e} feed & coud\'{e} spec.,A,TI5 & 30,000 \\
17 06 2004 & 53173 & KPNO coud\'{e} feed & coud\'{e} spec.,A,TI5 & 30,000 \\
11 06 2005 & 53532 & KPNO coud\'{e} feed & coud\'{e} spec.,A,TI5 & 30,000 \\
13 06 2005 & 53534 & KPNO coud\'{e} feed & coud\'{e} spec.,A,TI5 & 30,000 \\
14 06 2005 & 53535 & KPNO coud\'{e} feed & coud\'{e} spec.,A,TI5 & 30,000 \\
23 09 2006 & 54001 & KPNO coud\'{e} feed & coud\'{e} spec.,A,TI5 & 30,000 \\
28 09 2006 & 54006 & KPNO coud\'{e} feed & coud\'{e} spec.,A,TI5 & 30,000 \\
26 09 2007 & 54369 & KPNO coud\'{e} feed & coud\'{e} spec.,A,TI5 & 30,000 \\
04 11 2007 & 54408 & KPNO coud\'{e} feed & coud\'{e} spec.,A,TI5 & 30,000 \\
27 04 2008 & 54583 & KPNO coud\'{e} feed & coud\'{e} spec.,A,TI5 & 30,000 \\
07 05 2008\tablenotemark{b} & 54593 & TSU AST 2m & spec.,echelle,SITe ST-002A & 35,000 \\
29 04 2009 & 54950 & KPNO coud\'{e} feed & coud\'{e} spec.,A,TI5 & 30,000 \\
30 04 2009 & 54951 & KPNO coud\'{e} feed & coud\'{e} spec.,A,TI5 & 30,000 \\
20 06 2009 & 55002 & KPNO coud\'{e} feed & coud\'{e} spec.,A,TI5 & 30,000 \\
21 06 2009 & 55003 & KPNO coud\'{e} feed & coud\'{e} spec.,A,TI5 & 30,000 \\
27 04 2010 & 55313 & KPNO coud\'{e} feed & coud\'{e} spec.,A,TI5 & 30,000 \\
12 10 2011\tablenotemark{c} & 55846 & TSU AST 2m & spec.,echelle,Fairchild 486 &
25,000 \\
09 10 2012 & 56209 & KPNO coud\'{e} feed & coud\'{e} spec.,echelle,T2KB & 72,000 \\
20 04 2013 & 56402 & KPNO coud\'{e} feed & coud\'{e} spec.,A,STA3 & 26,600 \\
22 05 2013 & 56434 & KPNO coud\'{e} feed & coud\'{e} spec.,A,STA3 & 26,600 \\
26 10 2013 & 56591 & KPNO coud\'{e} feed & coud\'{e} spec.,echelle,T2KB & 72,000 \\
24 04 2014 & 56771 & KPNO coud\'{e} feed & coud\'{e} spec.,echelle,T2KB & 72,000 \\
30 07 2014 & 56868 & KPNO coud\'{e} feed & coud\'{e} spec.,echelle,T2KB & 72,000 \\
\enddata
\tablenotetext{a}{See 
Table~\ref{tbl-173093obs} 
for complete list of observing
dates, RVs, and sources.}
\tablenotetext{b}{First of 158 observations.}
\tablenotetext{c}{First of 133 observations.}
\end{deluxetable}

The KPNO CCD spectra were calibrated and extracted with standard 
IRAF tasks, after which radial velocities were measured with the 
IRAF task FXCOR \citep{fit93}. Template stars for the cross 
correlations were from the list of \citet{S10} or  
\citet{nid02}. A comparison of velocities from these two sources 
shows agreement within 0.1 km~s$^{-1}$ \citep{hor20}. 
Between 2008 May and 2020 June we extensively supplemented the 
KPNO spectra with an additional 291 usable spectroscopic 
observations acquired at Fairborn Observatory in southeast Arizona. 
The spectra were obtained with the Tennessee State University 2~m 
Astronomical Spectroscopic Telescope (AST) and fiber fed echelle 
spectrograph \citep{ew04}. Our initial detector was a SITe ST-002A
CCD with 15 $\mu$m pixels. In the summer of 2011 that CCD was 
replaced with a larger format Fairchild 486 CCD that also had 15
$\mu$m pixels \citep{fetal13}. Additional information is provided 
in Table~\ref{tbl-instcomb}. 
\citet{ew07} discussed the reduction and wavelength calibration of
the raw spectra.

\citet{fetal09} provided a general description of the velocity 
measurement procedure. In particular for HD~173093, we used a solar
line list that consists of 168 mostly neutral metallic lines in the 
spectral region 4920--7100~\AA\ and fitted each line with a rotational 
broadening function \citep{fg11,lf11}.

\begin{deluxetable}{lrrrrrrrrrrrrl}
\tabletypesize{\tiny}
\tablewidth{0pt}
\tablecolumns{11}
\tablenum{6}
\tablecaption{Radial Velocity Data\label{tbl-173093obs}}
\tablehead{
& 
\colhead{Measured} & \colhead{Model} & & &
\colhead{Measured} & \colhead{Model} &  & &
\colhead{Measured} & \colhead{Model} & &
& \\
\colhead{Modified} & 
\colhead{RV, Aa} & \colhead{RV, Aa} & O-C,Aa & \colhead{$\sigma_{\rm Aa}$} &
\colhead{RV, Ab} & \colhead{RV, Ab} & O-C,Ab & \colhead{$\sigma_{\rm Ab}$} &
\colhead{RV, B} & \colhead{RV, B} & O-C,B & \colhead{$\sigma_{\rm B}$} 
& \colhead{Set\tablenotemark{a}} \\
\colhead{Julian Date} & \colhead{(km/s)} & \colhead{(km/s)} & \colhead{(km/s)} & \colhead{(km/s)} &
\colhead{(km/s)} & \colhead{(km/s)} & \colhead{(km/s)} & \colhead{(km/s)} &
\colhead{(km/s)} & \colhead{(km/s)} & \colhead{(km/s)} & \colhead{(km/s)} &
}
\startdata
52073.4155 & ... & ... & ... & ... & -71.000 & -71.160 & 0.160 & 0.880 & ... & ... & ... & ... & 1\tablenotemark{b} \\
52074.4101 & ...& ... & ... & ... & ... & ... & ... & ... & -41.700 & -42.153 & 0.453 & 0.580 & 1 \\
52075.3773 & ... & ... & ... & ... & -52.100 & -50.012 & -2.088 & 0.880 & -41.100 & -42.134 & 1.034 & 0.580 & 1 \\
52402.4801 & -76.700 & -77.418 & 0.718 & 0.960 & -34.700 & -34.878 & 0.178 & 0.880  & -34.700 & -34.318 & -0.382 &0.580 &  1 \\
52403.4793 & ... & ... &. ... & ... & -72.300 & -73.018 & 0.718 & 0.880 & ... & ... & ... & ... & 1 \\
52545.1551 & ... & ... & ... & ... & -81.200 & -81.186 & -0.014 & 0.880 & ... & ... & ...& ... & 1 \\
52547.1323 & ... & ... & ... & ... & ... & ...& ... & ... & -28.900 & -29.586 & 0.686 & 0.580 & 1 \\
52556.0851 & -76.800 & -77.039 & 0.239 & 0.960 & ... & ... & ... & ... & ... & ... & ... & ... & 1 \\
52557.0857 & ... & ... & ... & ... & -81.100 & -81.894 & 0.794 & 0.880 & ... & ... & ... & ... &  1 \\
52558.0997 & -79.100 & -79.875 & 0.775 & 0.960 & ...& ... & ... & ... & ... & ... & ... & ... & 1 \\
52709.533 & -71.900 & -70.542 & -1.358 &1.004 & -49.700 & -50.028 & 0.328 &  0.679 & -24.600 & -25.340 & 0.740 & 1.518 & 2 \\
53172.3861 & -20.200 & -19.705 & -0.495 & 0.710 & -63.200 & -62.948 & -0.252 & 0.679 & -65.300 & -65.322 &  0.022 & 1.518 & 2 \\
53173.3669 & -55.200 &-56.348 & 1.148 & 1.004 & -24.400 & -24.302 & -0.098 & 0.679 & -64.300 & -65.310 & 1.010 & 1.518 & 2 \\
53532.4292 & -56.100 & -56.438 & 0.338 & 1.004 & -29.600 & -29.808 & 0.208 & 0.679 & -60.700 & -59.809 & -0.891 & 1.518 & 2 \\
53534.4298 & -66.800 & -65.146 & -1.654 & 1.004 & -20.800 & -20.652 & -0.148 & 0.679 & -61.400 & -59.778 & -1.622 & 1.518 & 2 \\
53535.4437 & -25.400 & -24.126 & -1.274 & 1.004 & -64.500 & -63.944 & -0.556 & 0.679 & -62.400 & -59.763 & -2.637 & 1.518 & 2 \\
54001.1377 & -66.400 & -65.699 & -0.701 & 1.004 & -27.700 & -26.927 & -0.773 & 0.679 & -55.700 & -53.043 & -2.657 & 1.518 & 2 \\
54006.1433 & -66.800 & -67.805 &  1.005 & 1.004 & -24.800 & -24.777 & -0.023 & 0.679 & -51.500 & -52.972 &  1.472 & 1.518 & 2 \\
54369.1752 & -71.200 & -71.135 & -0.065 & 1.004 & -27.900 & -26.634 & -1.266 & 0.679 & -49.300 & -47.699 & -1.601 & 1.518 & 2 \\
54408.0740 & -29.100 & -28.136 & -0.964 & 1.004 & -73.000 & -72.611 & -0.389 & 0.679 & -49.200 & -47.096 & -2.104 & 1.518 & 2 \\
54583.4562 & -66.600 & -66.203 & -0.397 & 1.004 & -34.900 & -35.393 &  0.493 & 0.679 & -45.700 & -44.207 & -1.493 & 1.518 & 2 \\
54593.4369 & -68.700 & -68.588 & -0.112 & 0.370 & -33.600 & -33.320 & -0.280 & 0.735 & -43.500 & -44.162 &  0.662 & 0.520 & 3 \\
\enddata

\tablenotetext{a}{Set 1 = KPNO velocities obtained by D. Willmarth, Set 2 = KPNO
velocities obtained by F. Fekel, Set 3 = Fairborn Observatory.}
\tablenotetext{b}{In this and in following lines of Set 1 data, three dots indicate
that the velocity for a given component was not used in the fit due to extensive line blending.}

\tablecomments{Table 6 is published in its entirety in the machine-readable format.
      A portion is shown here for guidance regarding its form and content.}
\end{deluxetable}

All of our spectroscopic observations and velocities for HD~173093 
(324 in total) are listed in 
Table~\ref{tbl-173093obs}. These show that HD~173093 is a triple 
system with the lines of all three components visible in the 
spectrum. The lines of the components can be distinguished from 
each other because of their different relative line depths and 
somewhat different widths. However, velocity measurement of the 
lines is more difficult than usual because the semi-amplitudes of 
the stars in the short-period system are not large enough to 
completely separate the three sets of lines at the resolutions of 
our various spectra. Therefore, in both the KPNO and AST spectra, 
the lines of at least two of the three components, the long-period 
component and one of the short-period components, are always at 
minimum partly blended. Thus, measurement of the blended lines 
consisted of simultaneous fits to the blended components. Our 
unpublished velocities for several IAU solar-type velocity 
standards show that our velocities with the SITe CCD have a 
$-$0.3 km~s$^{-1}$ shift relative to the results of \citet{S10}, 
and for the Fairchild CCD there is a $-$0.6 km~s$^{-1}$ shift. 
Thus, depending on the detector used, we have added 0.3~km~s$^{-1}$ 
or 0.6~km~s$^{-1}$ to the measured AST velocities.  

Our radial velocities from the two observatories are tied to the IAU
standards observed by \citet{S10}. Thus, there is no significant
zero point shift in the velocities from the two observatories
\citep{wetal16}.

\subsubsection{Combined Orbit}

Using both the spectroscopic data and speckle observations of HD 173093, we were able
to calculate a combined visual-spectroscopic orbit of this triple system using the methods
outlined in \citet{mut10}. Fitted velocities obtained from the model and observed minus
calculated ($O-C$) residuals for each component are also shown in Table~\ref{tbl-173093obs}.
The inner subsystem has a period of only 2.36 days, so it is 
inaccessible to our speckle observations. Because the inner orbit is so short, we 
tried the calculations two ways, first by fixing the eccentricity of the inner orbit to zero,
and next by allowing it to float. In either case, the elements of the outer orbit are unaffected, 
as are the derived masses of the three components and the distance to the system. For the purposes
of the discussion in the next section, we will work with the numbers obtained from the zero-eccentricity case.
(The eccentricity obtained in the second orbit was small but non-zero, $0.0056 \pm 0.0014$.)
Radial velocity plots for both orbits and the visual orbit of the wider component are shown in Figure 6, 
and the orbital elements, masses, and the distance derived 
are shown in Table 7. The distance we determine (75.79 $\pm$ 0.71 pc) is slightly larger than the 
value implied by the {\it Gaia} EDR3 parallax (72.56 $\pm$ 0.38 pc) while the Hipparcos revised value
\citep{van07} straddles both (73.9 $\pm$ 2.3 pc), although with lower precision. 
However, the basic astrometric information on the {\it Gaia} DR3 sources, obtained by treating 
all of them as single stars, has already been provided in EDR3 and will not change for DR3
\citep{lin20}.
As HD 173093 is unresolved in EDR3, this may be the source of the discrepancy.

\begin{deluxetable}{lrr}
\tabletypesize{\footnotesize}
\tablenum{7}
\tablewidth{0pt}
\tablecaption{Orbital Elements of HD 173093 = YSC 133  \label{tbl-orbelem}
}
\tablehead{\colhead{Parameter} & \colhead{Outer Orbit} & \colhead{Inner Orbit}
}
\startdata
$P$ (days)         & 2642.5 $\pm$ 1.5  &  2.3580103 $\pm$ 0.0000014 \\
$T$ (MJD)         & 55470.03 $\pm$ 0.63  &  55999.72733 $\pm$  0.00063 \\
$e$                    & 0.6109 $\pm$ 0.0013 &  0 (fixed) \\
$a$ (mas)          & 80.09 $\pm$ 0.99    &  ... \\
$i$ (deg)            & 104.55 $\pm$ 0.27 & 11.308 $\pm$ 0.032    \\
&& {\bf (or 168.692 $\pm$ 0.032)} \\
$\Omega$ (deg)  &  286.62 $\pm$ 0.66 & ... \\
$\omega{_A}$ (deg)  & 254.98 $\pm$ 0.17 & 0 (assigned) \\
$\mathcal{M}_{\rm A}$ ($M_{\odot}$)   & 2.857 $\pm$ 0.021 & ... \\
$\mathcal{M}_{\rm Aa}$ ($M_{\odot}$) & ... & 1.467 $\pm$ 0.011 \\
$\mathcal{M}_{\rm Ab}$ ($M_{\odot}$) & ...& 1.390 $\pm$ 0.011\\
$\mathcal{M}_{\rm B}$ ($M_{\odot}$)   & 1.416 $\pm$ 0.011 & ...  \\
Distance (pc) & 75.79 $\pm$ 0.71 &  ...  \\
\enddata
\end{deluxetable}

\begin{figure}[!b]
\figurenum{6}
\includegraphics[scale=0.37]{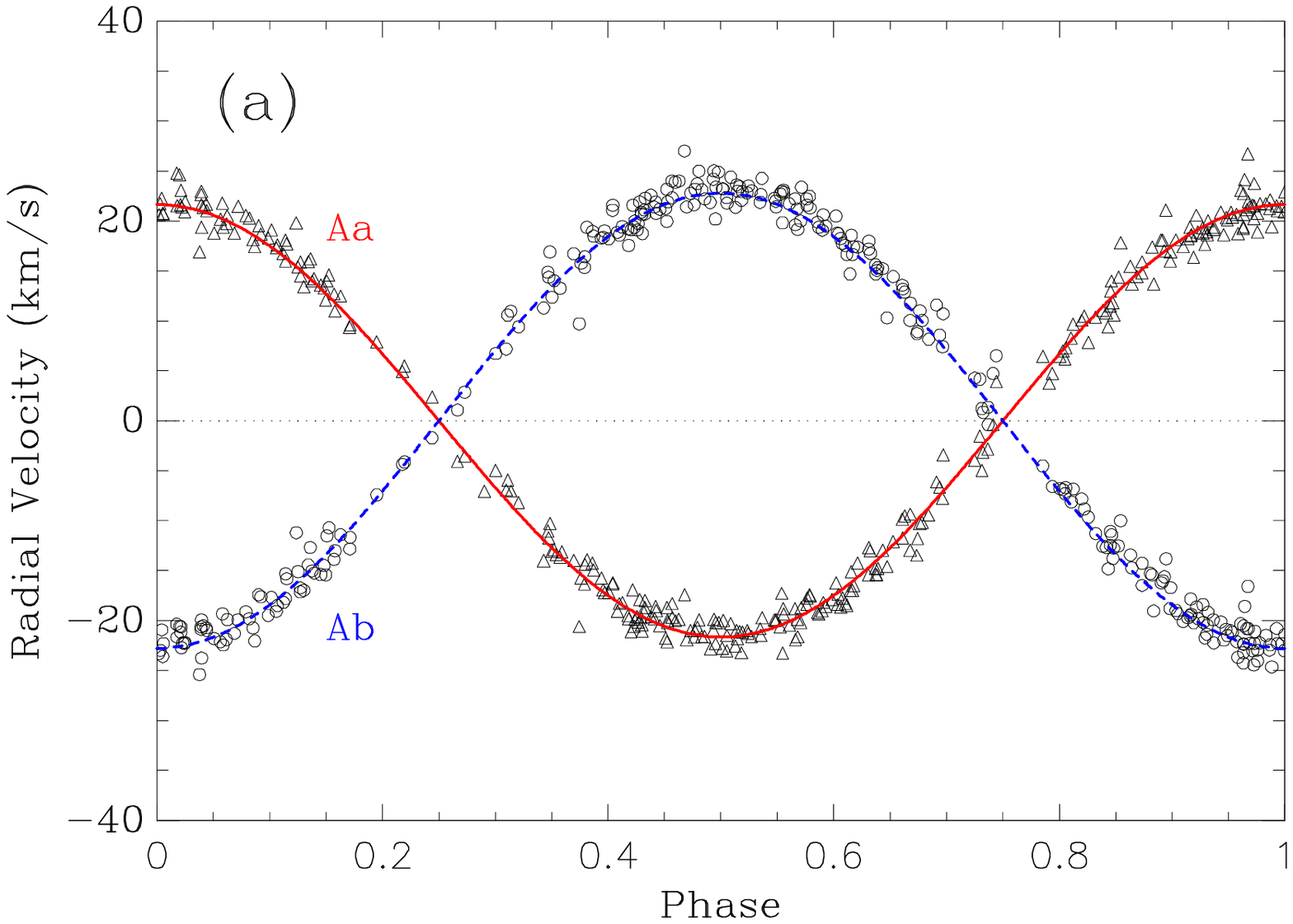}
\includegraphics[scale=0.37]{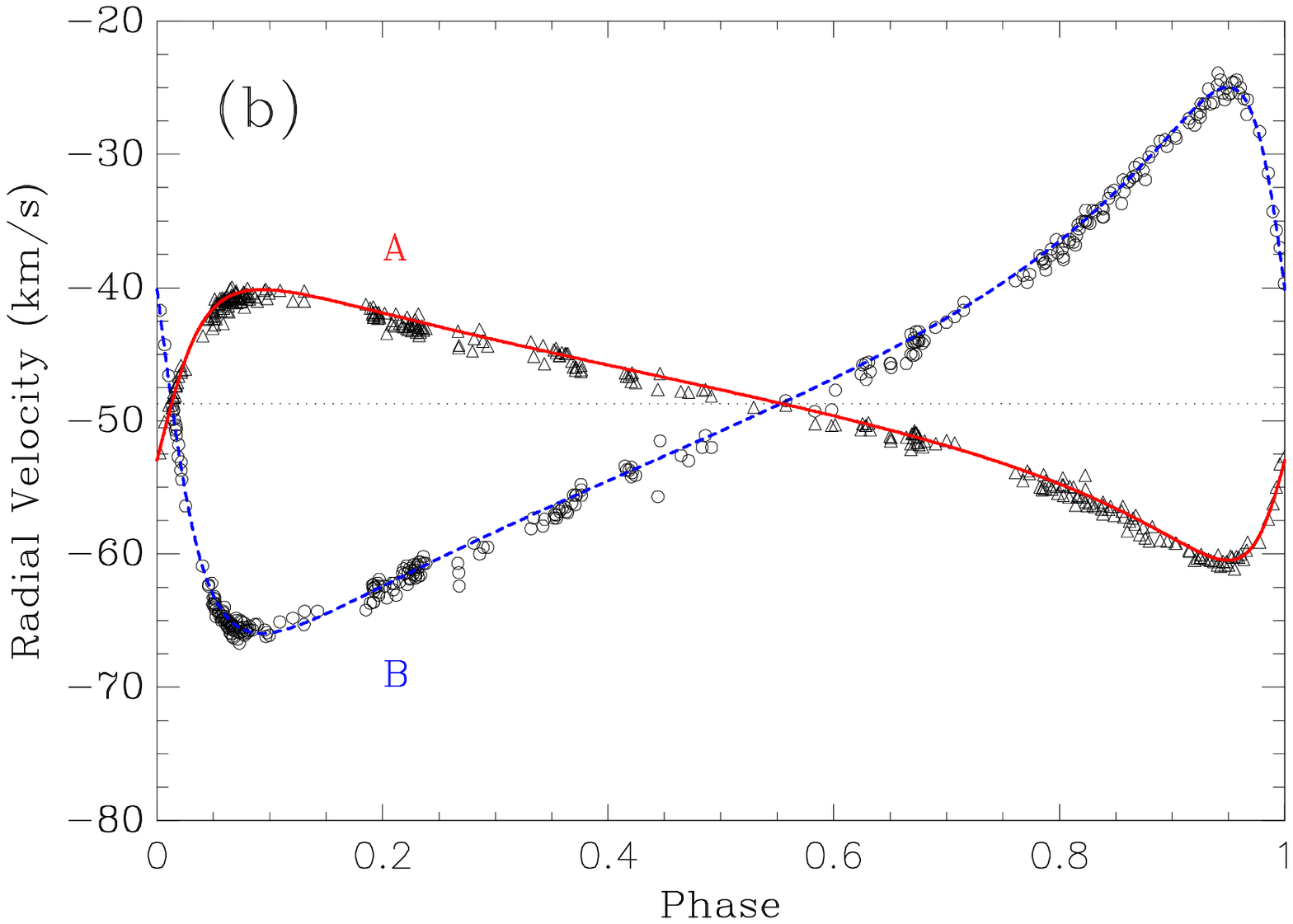}
\hspace{0.1cm}
\includegraphics[scale=0.37]{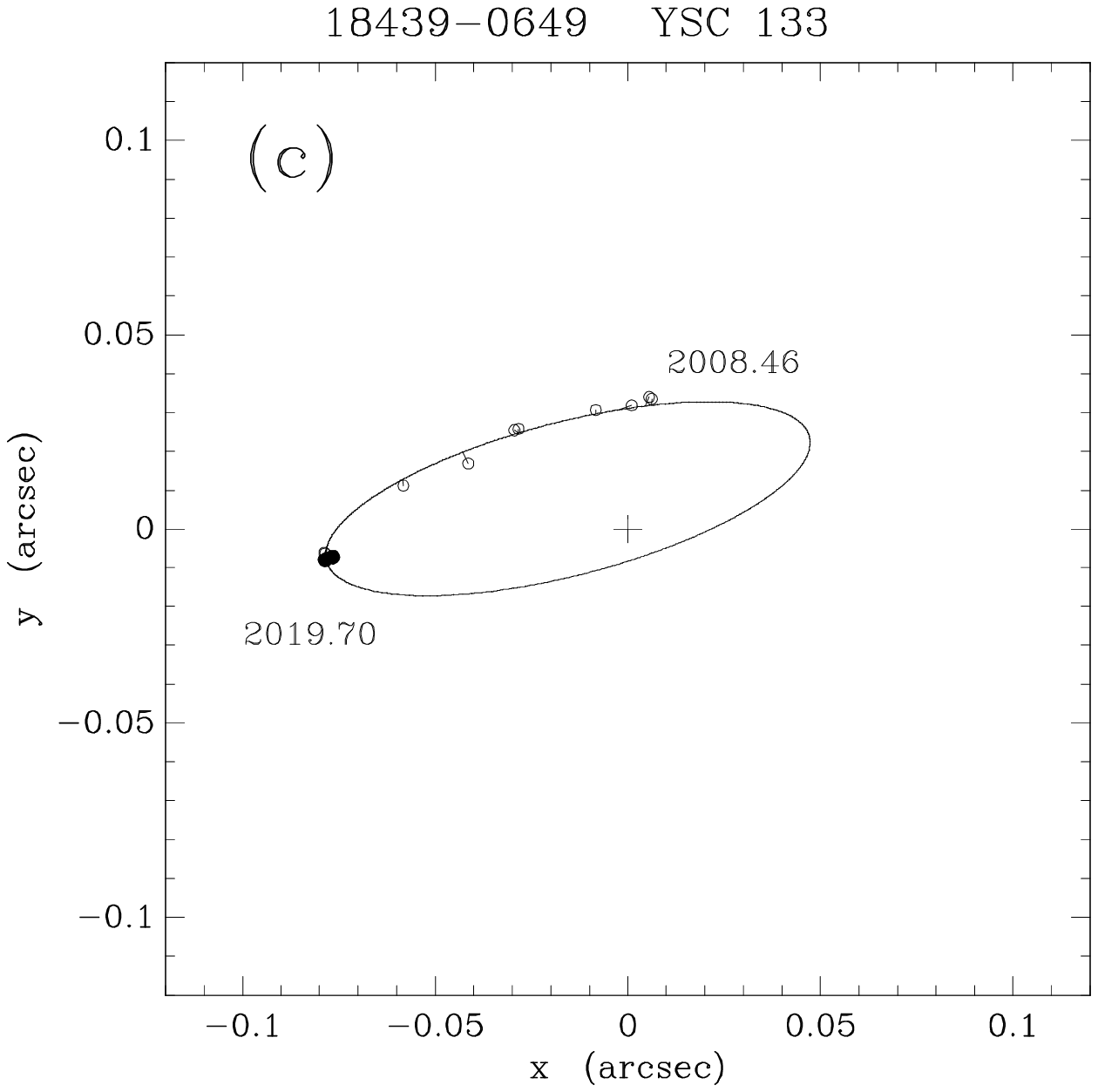}
\figcaption{(a) Radial velocity curves for HD 173093Aa and Ab, with the motion due to the third component subtracted. 
(b) Radial velocity curves of HD 173093 A and B.
(c) The visual orbit of YSC 133 = HD 173093AB as deduced from the combined astrometric and spectroscopic
data set. Here, open circles represent previous measures in the literature and appearing in the 4th Interferometric
Catalog, filled circles are data points appearing in Table 3, and line segments are drawn from the ephemeris position
on the orbit to each data point. The cross in the center marks the origin. North is down and east is to the right.}
\end{figure}

Based on our results, HD 173093 is a hierarchical triple system with a long-period to
short-period ratio of 1121. The inclinations of its two orbits
(Table 7) are very different, 104.6\arcdeg\ for the outer orbit and
11.3\arcdeg\ for the inner orbit, so the orbits are clearly not
coplanar. With such a large inclination difference, the eccentricity
and inclination of the inner orbit can undergo periodic changes
that are known as Kozai-Lidov cycles \citep{k62, l62}. According
to theoretical analyses, the evolution of such triples is driven
by the Kozai-Lidov modulation plus tidal friction. These processes
produce an inner binary with a period of just a few days
\citep{ms79, ek01, ft07}. In particular, \citet{ft07} predicted
that the combination of Kozai-Lidov cycles and tidal friction
typically results in inner binaries with periods less than 10
days. Their final short-period distribution had its peak at 3 days.
Observationally, \citet{tetal06} surveyed 165 solar-type binaries
and found that 96\% of those with periods less than 3 days were
triple. HD~173093, which has an inner orbit with a period of 2.36 days,
is consistent with the results of \citet{ft07} and \citet{tetal06}.

\subsection{Individual Masses for Six Stars}

The above orbital results afford us the opportunity to determine individual masses stars in the two 
triple systems discussed, HDS 1260 and HD 173093. 
Four of these masses are determined dynamically, whereas two rely on an assumption that those
stars have the same mass. In preparation for a comparison between the
masses and absolute magnitudes of these stars with the empirical data available in the most recent papers,
we have used the data tables in \citet{tor10} for F,G,K stars together with the polynomial fit derived in 
\citet{ben16}. Specifically, we derive the following polynomial fit based on the former reference for
$1 \leq M_{V} < 9$:

\beq
\mathcal{M} = 2.371 -0.2683 \cdot M_{V} -0.0483 \cdot M_{V}^{2} +0.0131 \cdot M_{V}^{3} 
-0.000761 \cdot M_{V}^4, 
\eeq

\noindent
where $\mathcal{M}$ represents the mass of the star. On the other hand, for $9 \leq M_{V} \leq 18$,
and defining $x = M_{V} - 13.0$, \citet{ben16} find that

\beq
\mathcal{M} = 0.19226 -0.050737 \cdot x + 0.010137 \cdot x^{2} -0.00075399 \cdot x^{3} 
-0.000019858 \cdot x^{4}.
\eeq

\noindent
Using these two formulas over the full range $1 \leq M_{V} \leq 18$ results in the plot shown in Figure 7,
which we will use to compare the masses we derive with previously known results. The data in 
\citet{tor10} also permits us to make a fit to the spectral type; although very approximate,
we use that fit as a basic guide in the discussion below.

For HIP 42910, the speckle measures are exclusively 
at red or near-infrared wavelengths, but the magnitude differences are near 3 for A-BC
and between 0.0 and 0.1 for the BC pair. {\it Hipparcos} gives a result for the magnitude 
difference in the $H_{p}$ filter,
which has has center wavelength of 511 nm with a width of 222 nm; that value is $3.53 \pm 0.27$. Using the apparent
magnitude of 10.16 from {\it Hipparcos} and the EDR3 parallax of $27.5131 \pm 0.0802$ mas, we find that
the absolute magnitudes of A and BC are likely near 7.40 and 10.90, respectively, where a $\Delta m$ 
at $V$ of 3.5 is assumed. Assuming a magnitude
difference of zero for BC, we can roughly assign an absolute magnitude to each those stars of 11.65. 

Calling upon Figure 7 to convert these absolute magnitudes to masses, we obtain 
individual masses and spectral types as follows:
$\mathcal{M}_{A} \approx 0.77 \msun$, a K6V star and 
$\mathcal{M}_{B}  = \mathcal{M}_{C} \approx 0.28 M_{\odot}$, putting both fainter stars near M3V.
Thus, from the photometry, we would conclude that the total mass of the triple is 
approximately 1.33 $M_{\odot}$.
The orbital information and the parallax allow us to determine that the total mass
based on the astrometry is $1.90 \pm 0.28 M_{\odot}$, and that the BC pair has total mass 
$0.91 \pm 0.16 M_{\odot}$. Subtracting these, $\mathcal{M}_{A} = 1.00 \pm 0.32 
M_{\odot}$, which agrees with
the photometric estimate within the uncertainty. 
Likewise if we assume the B and C components are in fact near-twins, then we can surmise that 
the individual mass in either of those cases is simply $\mathcal{M}_{B+C}/2 = 0.46 \pm 
0.08 M_{\odot}$.
However, in any case, both $\mathcal{M}_{A}$ and $\mathcal{M}_{B+C}$ are more massive than
would be expected based on photometry, and so further observations are warranted; this is a case where
very good masses for B and C are within reach in the coming years. The position of these stars on the 
MLR is shown in Figure 7. Finally, we note that the inclinations
of A-BC and BC indicate that the orbital plane of the two components are not aligned; this system may be a 
useful example in studies of star formation mechanisms as a result.

\begin{figure}[t]
\figurenum{7}
\hspace{4.5cm}
\includegraphics[scale=0.50]{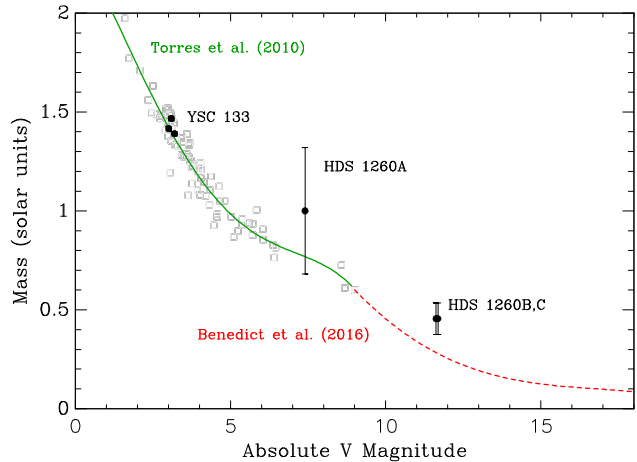}
\caption{
The empirical relation between absolute V magnitude and mass as described in the text. 
The open squares indicate the original data from \citet{tor10}, to give a sense of the 
observational scatter.
This relation is used to discuss the systems in this section. Positions of HDS 1260A, 
HDS 1260B and C (assuming they have equal mass as discussed in the
text), and all three components of HD 173093 (YSC 133) are also plotted with their
derived uncertainties in mass.}
\end{figure}

Turning now to HD 173093, which is another system with near-solar metallicity 
with [Fe/H] = -0.04 according to \citet{hol09}, we have a composite spectral type of 
F7V assigned by \citet{hou99}. Our orbital analysis in the previous section
reveals this to be a trio of F stars 
of very similar masses in a hierarchal arrangement. The magnitude difference from 
speckle observations for the AB component is well-determined from five measures
using a 543-nm filter, where $\Delta m = 0.62 \pm 0.07$. All measures in this case
are due to Tokovinin and his collaborators, e.g. \citet{tok16}. Using this value together
with estimates of the magnitude difference for the Aa,Ab component derived from the
spectroscopy, which is 0.1 magnitudes, we can deduce individual absolute V magnitudes
for the three stars to be 3.008, 3.092, and 3.192, respectively. Together with the masses,
these are used to locate the points on the mass-luminosity relation in Figure 7. We find
very good agreement with the curve fitting the \citet{tor10} data in all three cases.

\subsection{Notes on Four Other Systems}

Most of the visual orbits we have calculated are preliminary in nature and their main value at this stage is
in providing ephemerides for the coming years for observers to use as a guide while the orbits are 
further refined for the purpose of obtaining stellar masses. However, there are 
a handful of systems where a good fraction of the orbit is already traced out by the astrometric data at hand.
We give some further information on the systems appearing in Table 4 that have 
covered at least 270$^{\circ}$ in position angle in the time since the discovery observation. There are four
such systems, all with composite spectral types in the K range. These represent the orbits on
the firmest footing at present, and it is worthwhile in these cases to briefly examine the use of the
orbit in obtaining mass information. Plots of the orbits of these objects are shown in Figure 8.

\subsubsection{HIP 10616 = YSC 20}

YSC 20, a star suspected of being double based on the observations of the {\it Hipparcos} satellite \citep{esa97}, was first
resolved by our speckle program at the WIYN telescope in 2008 \citep{hor09}. SIMBAD shows the composite 
spectral type to be K0 and the {\it Gaia} EDR3 parallax value is $19.65 \pm 0.21$ mas. Given the apparent 
$V$ magnitude of 9.41 and using the magnitude difference of 2.63 at 562 nm in Table 3 as a proxy for
the $\Delta V$ of the pair, we estimate absolute magnitudes for the primary and secondary as 5.99 an 8.62,
respectively. Thus, this would appear to be a K0V (or very late G) primary and a late K secondary. 
Using Figure 7, we estimate masses of 0.88 and
0.66 $M_{\odot}$, meaning that $\mathcal{M}_{\rm tot} = 1.54 M_{\odot}$. 
On the other hand, the {\it Gaia} parallax
and the orbit we calculate here imply $\mathcal{M}_{\rm tot} = 1.41 \pm 0.07 M_{\odot}$.

\begin{figure}[!t]
\figurenum{8}
\plottwo{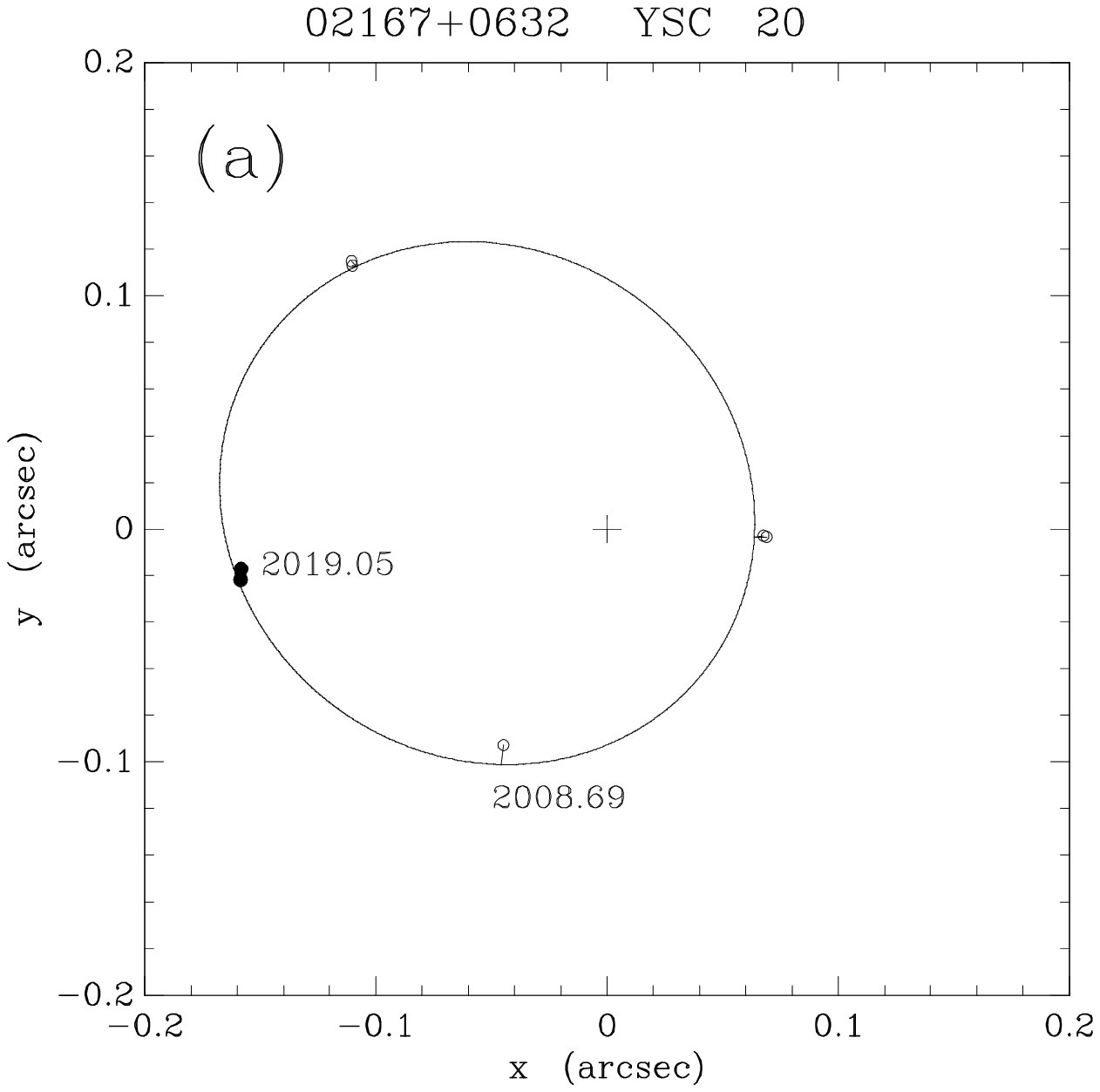}{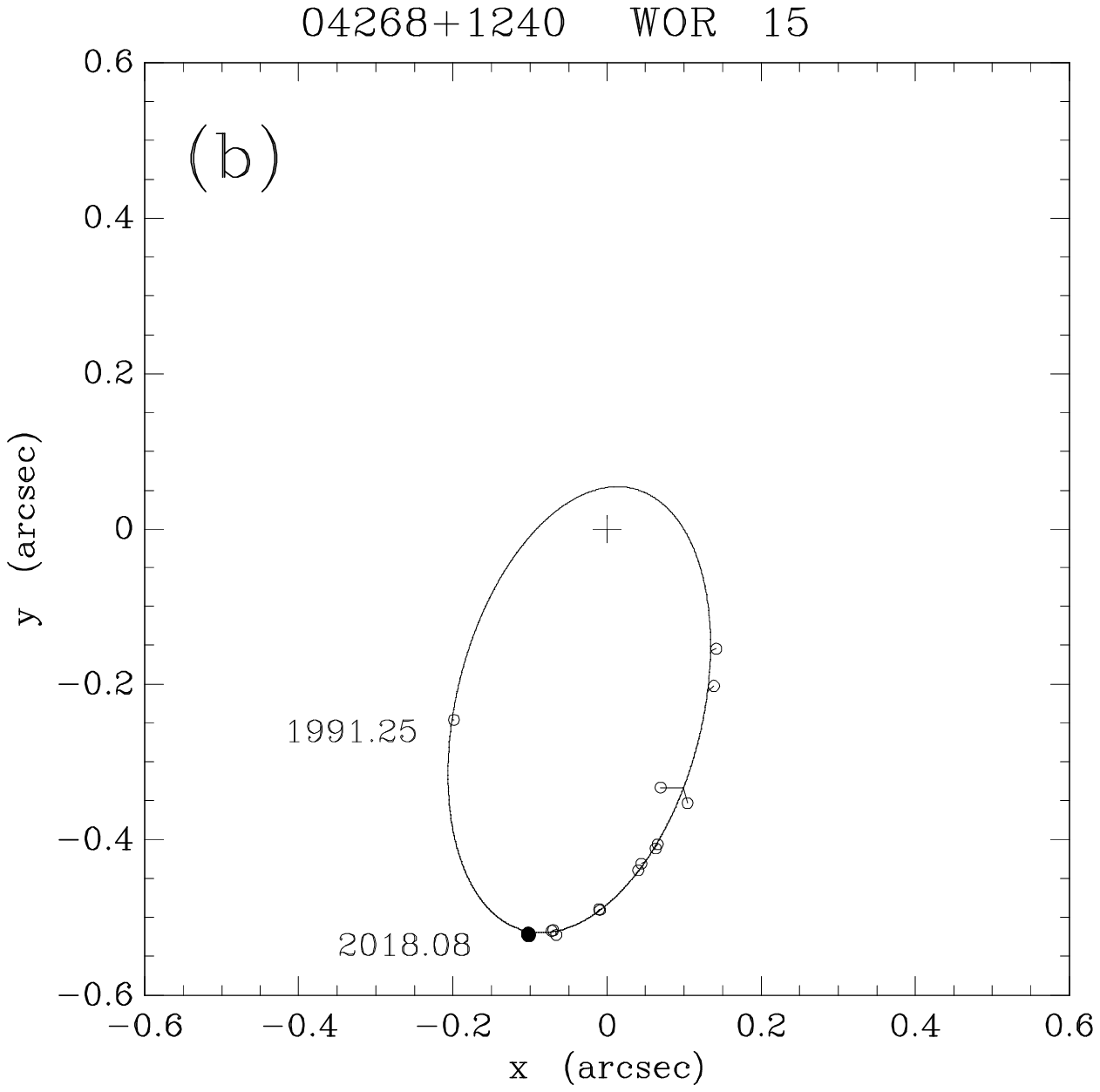}

\vspace{0.5cm}
\plottwo{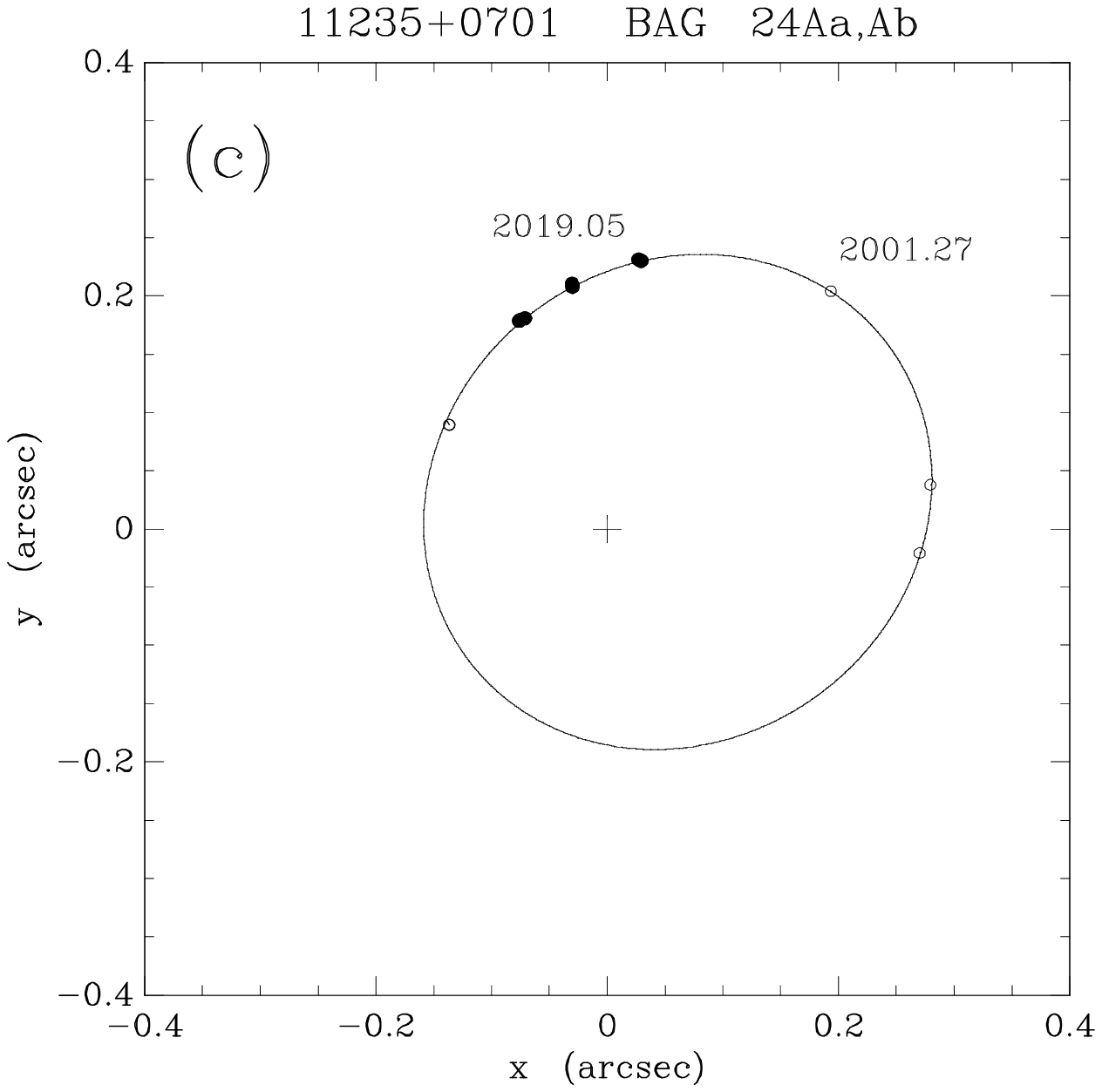}{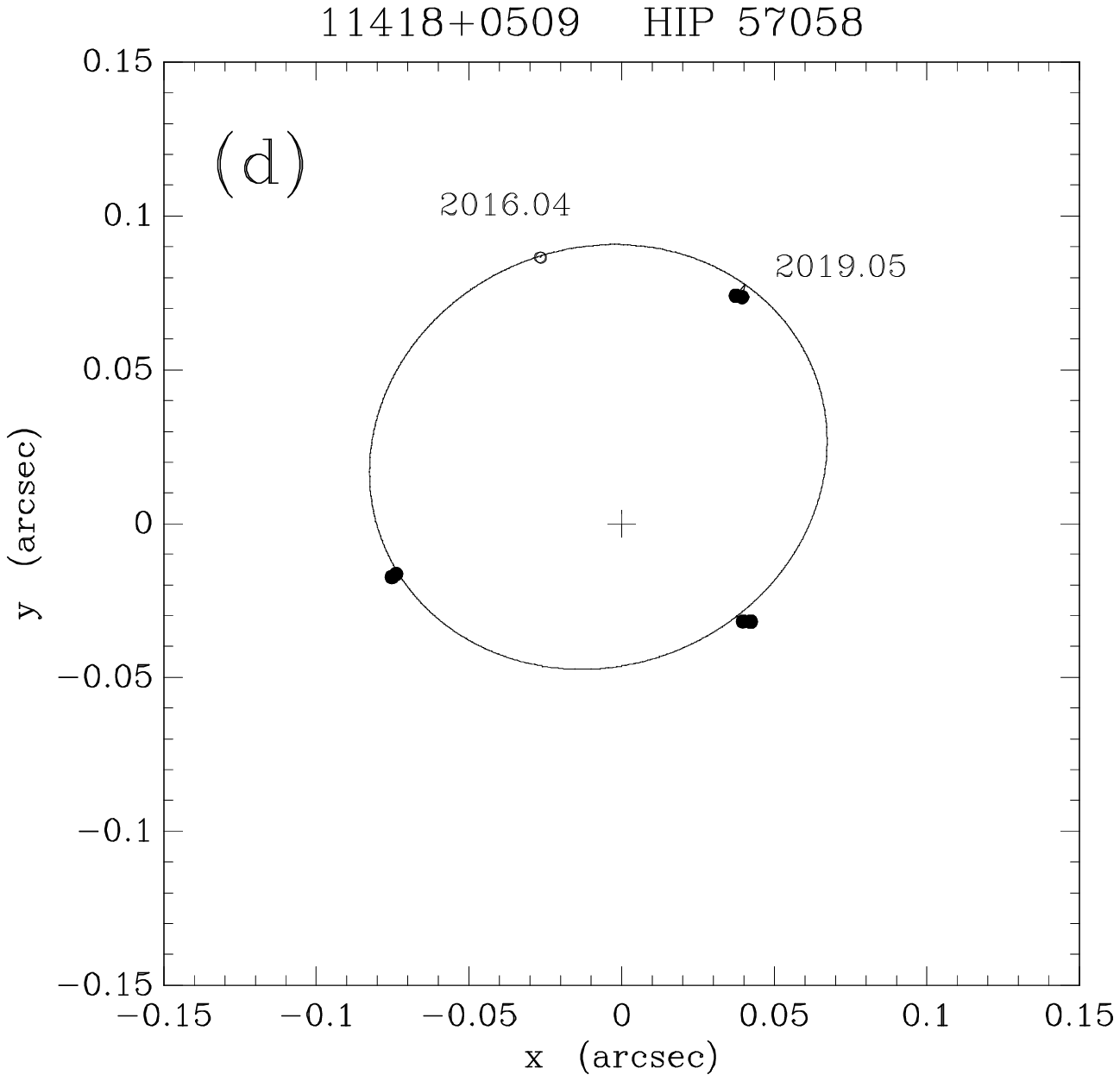}


\figcaption{Visual orbits in Table 4 for the four systems in Table 4 with the largest coverage
in position angle. (a) HIP 10616 = YSC 20; (b) HIP 20745 = WOR 15; (c) HIP 55605 = BAG 24Aa,Ab; 
(d) HIP 57058. In all plots, north is down and east is to the right. Measures appearing in Table 3 are shown
as filled circles while previous measures listed in the 4th Interferometric Catalog are drawn as open
circles. The cross marks the origin in each case and line segments are drawn from the ephemeris position
on the orbit to each data point.}
\end{figure}


\subsubsection{HIP 20745 = WOR 15}

Despite the spectral type of K2 shown in SIMBAD, photometry suggests that this pair
more likely consists of two mid- or late-K stars of near-equal brightness, with a $\Delta m$ of 0.25 based on two speckle
measures taken at 550 nm. The system is resolved in the {\it Gaia} EDR3 catalog (it was not in DR2, nor was
a parallax given in that release), and
the $\Delta G$ shown there is also small and consistent with the speckle result. Using that together with 
$V = 10.50$ and the EDR3 parallax of $24.1686 \pm 0.0878$ mas, we arrive at absolute $V$ magnitudes
of 8.05 and 8.30. Converting those values to masses, we obtain
0.72 and 0.70 $M_{\odot}$, and thus a total mass of 1.42 $M_{\odot}$. The orbital result is 
$1.32 \pm 0.62 M_{\odot}$, where the large uncertainty is due mainly to the fact that the semi-major
axis is not well-constrained by our orbit. Further observations near periastron, which will occur in the 2030's, 
will be important in calculating a definitive orbit of this system. We also note that the EDR3 catalogue 
gives parallax values for both components, and these indicate the two stars are at the same distance 
within the uncertainties.


\subsubsection{HIP 55605 = Bag 24Aa,Ab}

First reported in \citet{bag06}, this system has apparently gone unobserved for over a decade until 
the first of our observations shown in Table 3. From the 562-nm magnitude difference there and the
apparent magnitude and EDR3 parallax, we obtain absolute magnitudes of 7.69 and 9.65. These 
would suggest photometrically-determined masses of 0.75 and 0.50 $M_{\odot}$ 
using the same method as above, and 
thus a total mass of 1.25 solar masses. The orbit and distance give a total mass of $1.38 \pm 0.19 
M_{\odot}$, excellent agreement at this stage. As with the above system, a later spectral type is
implied from the photometry here than exists in SIMBAD (where it shows as K4V); we estimate K8V and M1V.

\subsubsection{HIP 57058}

Another pair with a composite spectral type of K4V according to SIMBAD, HIP 57058 was first resolved
by our speckle program in 2016 at Gemini-North \citep{nus21}. Since that point, we have seen the stars
complete nearly a full orbit in the subsequent observations shown in Table 3. Combining the known 
photometry with the EDR3 parallax result and converting to absolute magnitudes, the primary value 
is 8.06 and the secondary 8.55. These imply a total mass of 1.47 $M_{\odot}$, while the orbit and 
EDR3 parallax give $1.18 \pm 0.09 M_{\odot}$. 


\section{Basic Statistics of K-Dwarf Visual Binaries}

The orbits presented in the previous section consist largely of systems that have a K dwarf as the 
primary star. In this section, we add our subsample of these objects to the visual orbits 
previously known and available in the Sixth Orbit Catalog for this spectral type 
and make a preliminary study of the statistics of the sample. Confining our attention to K dwarfs within 50 pc of the 
Solar System, we find approximately 5000 such objects in current {\it Gaia} data. We have cross-identified these
objects with those in the Sixth Orbit Catalog, finding a grand total of 225 matches. Six of these objects
do not have a complete set of orbital elements, and so discarding those for our purposes here, we
are left with 219 previously known orbits. To those, we add the 21 K-dwarf orbits in Table 4, for a
grand total of 240 orbits to study. A period-eccentricity relation is shown for these objects in Figure 
9(a). It is worth noting that the majority of objects either have no stated uncertainty for eccentricity
or have $\delta e > 0.05$. Only 69 previous orbits have $\delta e \leq 0.05$, whereas for the
new orbits presented here, 16 systems meet that criterion. Therefore our new sample, shown with 
data points that are red squares, increases
this subsample by roughly 23\%. 

Two points are of note when studying the appearance of Figure 9(a). First, the period range represented
is clearly incomplete; nearly all the orbits plotted have periods between $10^{2}$ and $10^{6}$ days.
For a typical K-dwarf binary with total mass of 1 to 1.5$M_{\odot}$, these periods correlate
to semi-major axes in the range of 0.4 to 200 AU, or given the distances, typical angular values from 
$\sim$0.1 arcsecond to several arcseconds. That is, these are separations easily observable 
by direct imaging, speckle imaging, and adaptive optics at large telescopes. Outside of these period 
limits, we anticipate
that there are many more binaries; those with shorter periods are being found
in our spectroscopic program \citep{par21},
and those with larger periods would generally not have well-defined orbits due to the 
small amount of orbital coverage existing in the literature in most cases. 

Second, it is interesting to observe that
only three systems have an eccentricity less than 0.1 with estimated uncertainty below 0.05, which
are HIP 5842, HIP 34025, and HIP 78842, in order of increasing period. The orbit for HIP 5842 is 
the CD pair of a multiple system (I 27 CD), and the AB pair (HJ 3423AB) has a composite spectral 
type that indicates that
the primary has evolved off the Main Sequence. Thus, it may not be a particularly good system with 
which to judge K-dwarf orbital statistics. HIP 34025 (A 1959) has an orbit calculated by \citet{doc06} and only
one measure since that time appears in the 4th Interferometric Catalog. Docobo et al.\ find an
orbital period of 32 years based on five measures in the 4th quadrant and one (at the time) in the
second quadrant. However, the system has a small magnitude difference, and so quadrant 
determinations should be regarded as provisional. If instead all observations
are assumed to be in the same quadrant, then an orbit can be obtained with a 15.5-year period that 
appears to fit the data as well as Docobo's, but has an eccentricity of 0.76. Until further observations
are made of this system, it may be wise to assume that the eccentricity is not as certain as it 
appears at present. Finally, HIP 78842 (SEE 264AB) is also a quadruple system, 
with the orbit here representing the middle component in terms of separation;
the B component was first resolved in 2008 \citep{tok10} and is now known as 
WSI  84Ba,Bb, and a 10-arcsecond companion (SEE 264AC) is also present.
Orbital dynamics that cause the evolution of orbital elements over time may be at play here as well.
A group of only four other objects have periods near 1000 days and 
eccentricities below 0.2, though with substantial
uncertainty. These are HIP 1768, 4365, 16192, and 19832. All four are astrometric orbits due to 
\citet{gol07}, derived from very small photocenter shifts in {\it Hipparcos} data. Eccentricity values 
above 0.1-0.2 cannot be ruled out in any of these cases.
Perhaps {\it Gaia} can add to our knowledge here with future data releases, but the 
available data at present suggest the
relative lack of low-eccentricity systems, particularly for periods from at least $10^{2}$ to $10^{4}$ days. 

Figure 9(b) shows a histogram of the same samples as a function of log(period). Here we see that 
the orbits presented in Table 4 mainly contribute to the period range of $10^{4} - 10^{5}$ days, resulting
in a more strongly peaked histogram in that region for systems with well-measured eccentricities. 
The plot is of course incomplete but points to the possibility of significant improvement through sustained 
spectroscopic and speckle observing in the coming years, as is our long term goal.

\begin{figure}[!t]
\figurenum{9}
\plottwo{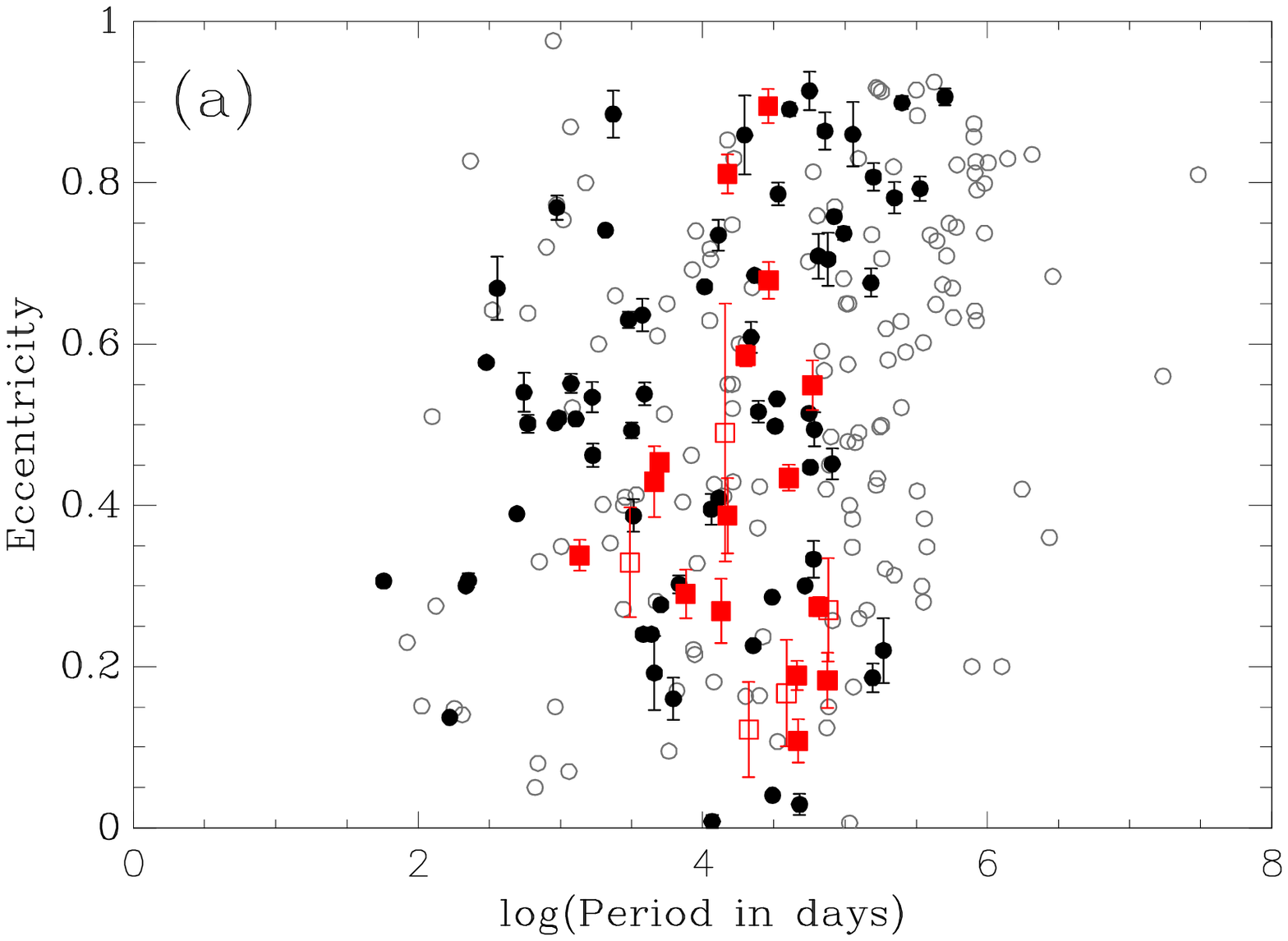}{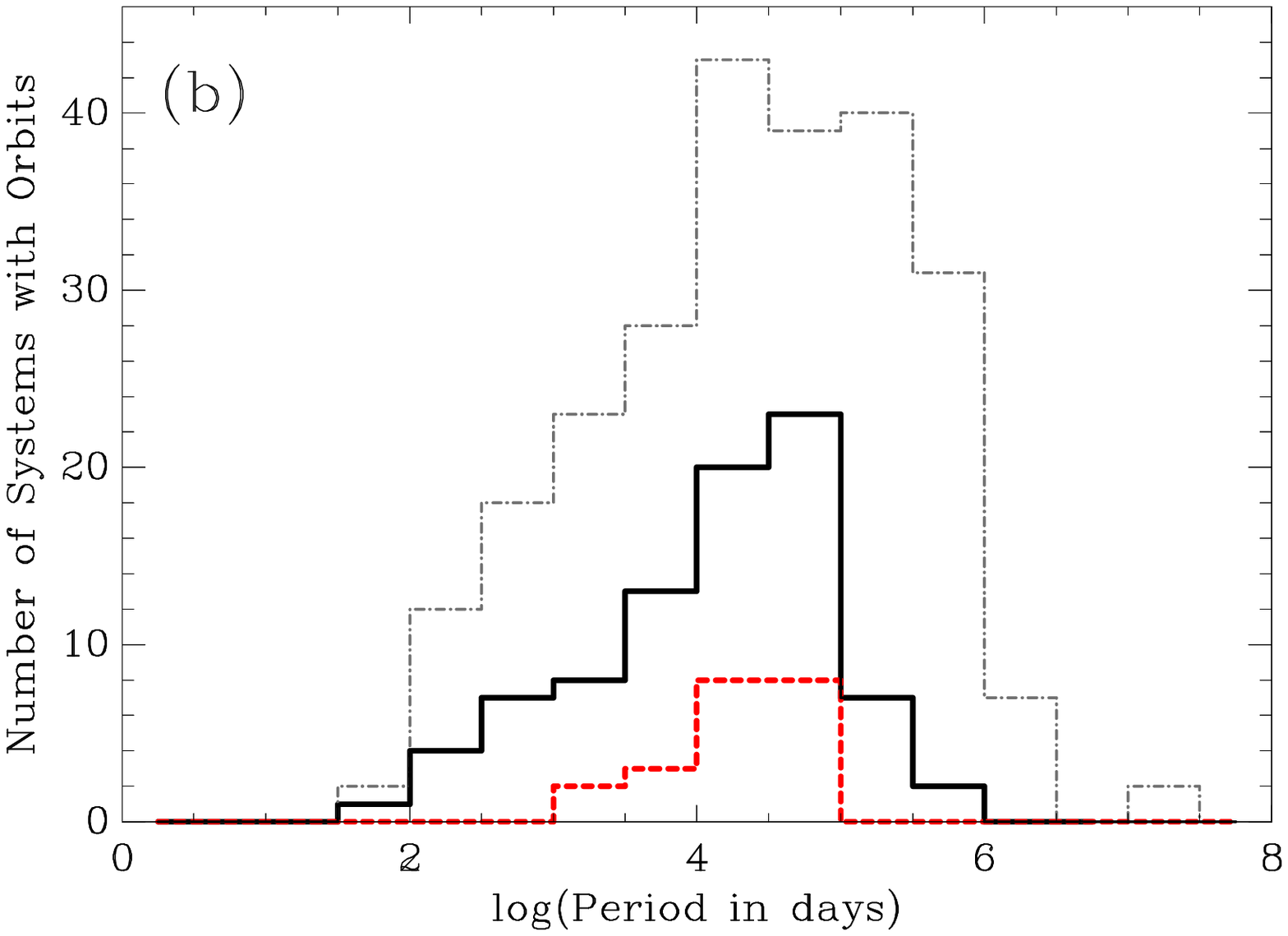}
\figcaption{(a) Period-eccentricity diagram for the K stars with known orbits. The orbits in the Sixth
Orbit Catalog are shown with circles, and the orbits presented in Table 4 are represented as red squares.
If an orbit has an uncertainty in eccentricity of greater than 0.05, then it is shown as an open grey or red
symbol. Uncertainties for published orbits in this case are not shown to keep the plot clear.
(b) Period histograms for three samples. Here, the dot-dashed grey histogram corresponds to 
all systems, the black data set represents those systems with $\delta e < 0.05$, and the red, 
dashed histogram indicates the systems in Table 4.}
\end{figure}

Finally, in Figure 10, we show the distance histogram of the same three samples: all K-dwarf orbits, 
those with well-known eccentricities, and our new orbits. If one assumes (1) the RECONS K dwarf 
number density of 0.01 per cubic parsec\footnote{\tt http://www.recons.org/census.posted.htm} 
and (2) multiplicity rate for K dwarfs that is between that for G and M stars,
we can begin to estimate how incomplete the current sample
is. As discussed above, the sample of visual orbits within 50 pc is mainly limited to the period 
range of $10^{2}$ to $10^{6}$ days, and if one examines the period histogram found in \citet{duq91}
then for G dwarfs we find that about half of all objects in their sample have periods falling in this range.
So, if we also assume a similar number for K dwarfs and a multiplicity rate like that of 
G dwarfs, we would predict the number of multiple 
star systems to follow the green curve in Figure 10. On the other hand, if the rate for M dwarfs 
is assumed, the dark red curve is obtained.
We see from this that the histogram of all 
known orbits is is therefore probably complete or nearly complete to about 20 pc, but appears very incomplete at larger distances. 
A much more robust sample can be made by doing two things. 
First, it will be important to increase the precision on the known orbital elements
for nearby systems paying particular attention to the eccentricity. This would move the black histogram
in Figure 10 closer to the grey one. Second, a full reconnaissance of systems 
beyond 20 pc is needed in order to eventually add many more orbits to the 25 we have presented here. 
(As seen with the red curve in Figure 10, all of the systems presented here are increasing the previously
undetermined sample of orbits for K dwarfs from 25--50 pc.)
Once these tasks are complete, the statistics of orbital elements of K dwarfs can be fully understood and 
a meaningful comparison to those with other spectral types can be made.

\begin{figure}[!htbp]
\figurenum{10}
\hspace{4.5cm}
\includegraphics[scale=0.50]{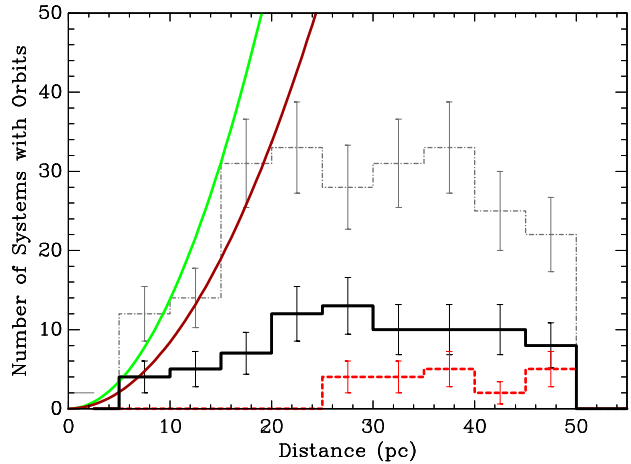}

\caption{
Distance histograms for the full sample of K dwarfs with orbits (shown in the dot-dashed grey), those
orbits with eccentricities with uncertainties of less than 0.05 (represented by the black histogram),
and the sample in Table 4 (in dashed red). 
The green and dark red curves represent the expected number of K dwarf binaries as a function of distance, making 
an assumption about the multiplicity rate in each case. For the green curve, the value of 44\% is used, which is
the known rate for G stars from \citet{rag10}, and for the red curve, we use 26.8\%, which is the rate for M
dwarfs from \citet{win19}.}
\end{figure}

\section{Conclusions}

This paper has first provided 378 new speckle observations of 178 binary star systems,
taken mainly with the LDT but supplemented with data from WIYN and both Gemini
telescopes. We have studied the astrometric and photometric precision of those measures, 
and conclude that they are in line with previous papers in this series. In particular, the 
estimated uncertainty for the separation measures is $2.07 \pm 0.11$ mas, and for the 
position angle it is of course dependent on the separation, but an average value for the
group of measures here is approximately 1.1 degrees. Photometric precision appears to be
on the order of 0.1 magnitudes, as expected for this sample, which mainly consists of stars 
brighter than $V = 12$. 

We used these measures and others in the literature to compute preliminary visual orbits of 25 
systems, 21 of which have a K dwarf as the primary star. 
Two triple systems provided information for mass determinations.  First, we
     had sufficient spectroscopic and speckle observations to compute
     combined orbits for HD 173093, yielding three high-precision masses of
     similar F dwarfs of 1.39--1.47 \msun.  Second, in the triple system HDS 1260 (HIP
     42910), we used the speckle results to determine component masses of
     1.00, 0.46, and 0.46 \msun, where these values are 
     larger than anticipated from photometry, indicating that more
     observations are warranted.

Finally, having identified a sample of 5000 K dwarfs within 50 pc of the Solar System using {\it Gaia} data, 
we cross-identified these with existing orbits in the Sixth Orbit Catalog and then added our new 
orbits to the sample. The period-eccentricity relation for these objects suggests a relative lack
of low-eccentricity orbits with period from 100 to 10,000 days, that is, a range comparable to those of
the planets in our own Solar System.

\acknowledgements 

E.P.H. gratefully acknowledges support from the National Science Foundation, specifically grants 
AST-1517824,  AST-1616698, and AST-1909560, which allowed all of the SCSU personnel to participate in this
work. Likewise, T.J.H. is grateful for NSF grants AST-1517413 and AST-1910130, and G.T.vB. acknowledges support from
NSF grant AST-1616084.
J.G.W. is supported by a grant from the John Templeton Foundation. The opinions expressed in this 
publication are those of the authors and do not necessarily reflect the views of the John Templeton 
Foundation. 
We thank Paul Klaucke, Richard Pellegrino, and Daniel Nusdeo for their help in obtaining some of the LDT observations, 
and members of the NASA Ames speckle group 
for participating in the Gemini observations discussed here.
Finally, we acknowledge the contribution of Helmut A. Abt in obtaining the initial
spectroscopic observations of HD 173093. Astronomy at Tennessee State University is supported by the 
state of Tennessee through its Centers of Excellence program.

This work was based in part on observations obtained at the Gemini Observatory, which is operated by the 
Association of Universities for Research in Astronomy, Inc., under a cooperative agreement with the NSF on 
behalf of the Gemini partnership: the National Science Foundation (United States), National Research Council 
(Canada), CONICYT (Chile), Ministerio de Ciencia, Tecnolog\'{i}a e Innovaci\'{o}n Productiva (Argentina), 
Minist\'{e}rio da Ci\^{e}ncia, Tecnologia e Inova\c{c}\~{a}o (Brazil), and Korea Astronomy and Space Science 
Institute (Republic of Korea). As visiting astronomers to Gemini-N, we are mindful that Maunakea is a sacred 
space to many native Hawai`ians, and we are grateful for the opportunity to have been present there. 

Some of the observations in the paper made use of the NN-EXPLORE Exoplanet and Stellar Speckle Imager 
(NESSI). NESSI was funded by the NASA Exoplanet Exploration Program and the NASA Ames Research 
Center. NESSI was built at the Ames Research Center by Steve B. Howell, Nic Scott, Elliott P. Horch, and 
Emmett Quigley. Kitt Peak is home to the Tohono O'odham people, and we are privileged to have been visiting
astronomers to that special place.






\end{document}